\documentclass[conference]{IEEEtran}

\IEEEoverridecommandlockouts

\usepackage[utf8]{inputenc}
\usepackage{cite}
\usepackage{amsmath,amssymb,amsfonts}
\usepackage{accents}
\usepackage{algorithmic}
\usepackage{graphicx}
\usepackage{textcomp}
\usepackage{xcolor}

\usepackage{empheq}

\usepackage{hyperref}

\usepackage{amsthm}
\theoremstyle{plain}

\usepackage{diagbox}

\newcommand*{\boldone}{\text{\usefont{U}{bbold}{m}{n}1}}

\newcommand*{\boldzero}{\text{\usefont{U}{bbold}{m}{n}0}}

\mathchardef\mhyphen="2D

\DeclareMathOperator*{\argmin}{argmin}

\usepackage{amsmath}
\newcommand{\thickhat}[1]{\mathbf{\hat{\text{$#1$}}}}

\renewcommand\footnoterule{\kern-3pt \hrule width 1in \kern 2.6pt}

\providecommand{\keywords}[1]
{
  \small	
  \normalfont{\textit{Keywords---}} #1
}

\title{A Catalog of Probability  Generating  Functionals 
for Multitarget Tracking and Data Assignment Problems  
}
\author{Roy L. Streit \\
\small Metron, 1818 Library St., Suite 600,  Reston, VA 20190  USA\\
\small 
r.streit@ieee.org   \emph{and}  streit@metsci.com} 

\usepackage{fancyhdr}
\usepackage{lastpage}
\begin{document}

\maketitle


\newpage

{\color{black}
\abstract This paper studies the class of Bayesian tracking filters for which the probability generating functional  of the joint target-measurement process can be
derived from  the  statistical assumptions that define the  problem. The class  includes filters for labeled  and unlabeled targets, as well as hybrid filters  in which  both labeled and unlabeled targets are present.  
New results include  the probability generating functional for interval filtering of target trajectories for both labeled an unlabeled targets,  and a novel method for computing the probability generating function  of measurement-to-target assignment   probabilities.  Probability generating functionals  give exact expressions for calculating  the importance  weights in particle  filter implementations. 
Low computational complexity  approximations to the particle weights are
derived from the  probability generating functionals via the saddle point method.         
}

\bigskip


\bigskip
{\color{black}
\section{Introduction}
\label{Introduction}}
 
This largely theoretical paper is devoted exclusively to the study of a specific class of Bayesian tracking problems, namely, those for which the \underline{p}robability \underline{g}enerating \underline{f}unctiona\underline{{l}} (PGFL) of the joint target-measurement process is derived   from  fundamental assumptions.  This class is large and diverse.  It comprises  probabilistic data association (PDA) filters  for labeled targets,  intensity filters  for unlabeled targets, and  hybrid filters for problems with both labeled and unlabeled targets.  These filters differ significantly from each other, yet they can all be  formulated exactly via their PGF{L}s.  The similarities and differences between them are reflected   in  their PGF{L}s in a way that words do not  convey as effectively.  

This paper is a catalog of tracking filters as seen through the lens of PGFLs. It is a comprehensive catalog of the PGFLs   that define the fundamental probabilistic and combinatorial structures of  many multisensor multitarget tracking filters and  the probability  generating functions  (PGFs) that characterize assignment problems. It is perhaps the first paper to adopt this  point of view exclusively. 

We demonstrate by example that PGFLs are elegant, concise, and complete models of the fundamental combinatorial assumptions that characterize  many tracking filters. This  evidence  supports the claim that PGFLs are a strong tool for posing and solving tracking problems. 

\subsubsection*{Contributions}The lens of PGFLs enables us to organize and derive a great many tracking filters from a single unified point of view. This  new perspective is  worthwhile  pedagogically, and it  enables us  to make new contributions to the tracking literature. Several are mentioned here. 

We derive   PGFs for  measurement assignment problems  in Section \ref{GFforAssignments}. These PGFs appear to be new, although the assignment problems themselves are not.  One example is the PGF of multisensor measurement-to-target assignment probabilities. Another   derives the PGF for measurement assignments in the  joint probabilistic data association (J-PDA) filter. A special case of the J-PDA result yields a new derivation of  an   elegant expression for an assignment probability as a ratio of matrix permanents---a  ratio that is shown to be the ratio of mixed first-order derivatives of the PGF. This expression  has been independently discovered  several times (see last paragraph of Section \ref{labeledJPDA}). The PGF derivative form of the assignment probability  is general, so the new derivation shows that the matrix permanent expression is not an isolated result.  

A  target trajectory process  is modeled as a state-dependent time-inhomogeneous branching process with ${K\geqslant 1}$ discrete time steps  and is initiated by a single target. The PGFL of this  process is derived in Section \ref{MultipleScan}.  It is the building block from which we derive  PGFLs for multitarget tracking filters for both labeled and unlabeled target trajectories.  This may be  the first application of PGFLs to state-dependent time-inhomogeneous branching processes models for  multitarget trajectory tracking. Many kinds  of branching processes have been studied  in depth over the last  century, with applications in biology (e.g., population dynamics) and physics (e.g., neutrons in a chain reaction), but this may also be the first appearance of this particular kind of branching process in the larger literature. The posterior PGF of  the number of target trajectories at any time in the interval is derived from the PGFL of the trajectory process via a marginalization procedure.  The derivation shows that  the  classical  state-dependent  Galton-Watson process is  the prior for the number of trajectories.

We derive  PGFLs for  several newly formulated tracking   problems that  will be the subject of future work but are  not explored further in this paper. The example  in Section \ref{DoublePoisson} derives  the PGFL for the problem of simultaneously localizing unknown numbers of both targets and sensors using ``pooled'' (unlabeled) sensor measurements.     The example   in Section \ref{toiJIPDAinaCloud} derives the PGFL for tracking a single  target-of-interest in a heterogeneous swarm of known and unknown targets that are not  individually tracked.

The exact posterior probability distribution of  every  PGFL-based tracking filter can be derived by computing the mixed first-order derivative of its  PGFL.  In virtually all cases, however, the Bayesian  recursion cannot be closed because the  exact posterior is not of the same mathematical form as the prior.  To close the recursion and produce an implementable  filter, it is necessary to approximate the exact posterior with a  distribution  of the same form as the prior distribution.  

A variety of problem-specific approximations to the posterior are available in the literature but are outside the scope of this paper, with the one exception.   The exception is discussed in Section \ref{saddlept}.  It is  an established method in the physics community known as the   saddle point method. It is a way to approximate  mixed derivatives of PGFLs without  differentiating them symbolically.     It is effective when (as is often the case in tracking) it is  easier to compute the PGFL  than to compute mixed first-order derivatives.   We illustrate the  method with the J-PDA particle filter.  Computing  exact particle weights  is an NP-hard problem, but it is seen that the complexity of the saddle point approximation of a particle   weight is at most $O(N^{2.4})$, where $N$ is the number of targets.  



\subsubsection*{Prior Work} ``Drawing principally on the physical and ecological contexts,'' the fundamental paper \cite{1962aMoyal} clearly defined the  mathematical constructs of PGFLs for the first time. It proved, e.g., that the derivatives of the PGFL generate  all the existing moment distributions of the process.  The paper \cite{1962bMoyal}  developed the theory  PGFLs for general branching processes, which are closely related to the trajectory processes discussed in Section \ref{MultipleScan}.    The beginnings of the  theory of PGFLs appeared in the physics community in 1935 and 1946 \cite{2003DaleyVereJones}.

This paper is tightly focused  on the PGFLs that characterize tracking filters, and not on filter implementation and  performance. There is little  work that prioritizes PGFLs in this way.   The first  use of PGFLs in tracking applications  appeared in 1976.  This work \cite{BakIvan} was published in English, and presumably also in Russian.  In it the authors derived the PGFL of what is now called the cardinalized probability hypothesis density (CPHD) filter and, as a special case, the PHD filter.  It  was one of a number of papers by multiple authors who worked in a field they called   ``stochastic flow'' and for whom the PGFL was a central tool. The work of several  authors, much of it published  in English at the time,    is reviewed in \cite{stochasticFlows}.  

The PHD  filter was  derived in 2003 using what are called random finite sets (RFSs) \cite{2003Mahler}, a concept  the author was careful to distinguish from  finite point processes. The PGFL derivation appeared in an appendix of that paper. The CPHD filter \cite{2007Mahler} appeared  in 2007. Extensions of the work using PGFLs include \cite{2012ClarkMahler} and \cite{2018ClarkMelo}.  A number of other filters by multiple authors were subsequently derived using RFSs, but most of these papers eschew the use of PGFLs. The PGFL formulations of several classes of multiBernoulli 
filters are given in \cite{2015Williams} and \cite{2016Mahler}.  The papers \cite{2015Williams} and \cite{2017BrekkeChitre} are   distinguished in that they show close  relationships to other tracking filters, specifically, the multiple hypothesis tracking (MHT) filter \cite{1979Reid} and joint integrated PDA (J-IPDA) filter \cite{2004MusickiEvans}.  The paper \cite{2017BrekkeChitre} extends results  in \cite{2015Williams} to derive the original MHT recursive update formulae.   

The classical PDA  filter was derived by the PGFL method in   \cite{2014bStreit}.   That paper initiated a series of subsequent papers  that are cited in the appropriate places in the present paper. Much of the work is included in \cite{2021ACBook}.        
   
Selected references to the larger (i.e., non-PGFL) filtering literature are, for  the most part,  limited to  early papers in which the filters first appeared in the literature.  These references are woefully incomplete, but they are  hopefully  sufficient to guide readers seeking to find the current tracking literature, which is very large.  Oversights and omissions are unintentional.

\bigskip
\section{Organization of the paper}
\label{organize}

To make the paper accessible to a larger audience, the Appendix provides an informal, but   terse, introduction to  PGFs and PGFLs for i.i.d.  (independent, identically distributed) finite point processes and bivariate i.i.d. cluster processes. The presentation is detailed enough so that readers can  calculate the PGFLs of  posterior point processes of the kind that are needed  for tracking filters. 

We divide the broad and diverse class of tracking problems that are  characterized by PGFs and PGFLs into two large subclasses---those that treat targets as identified and those that do not. Both divide further into single and multiple sensor versions, and both have interval (batch) filter generalizations.  The topics covered  are organized  as follows:     
\begin{itemize}
    \item \underline{Section \ref{AssocAndNotation}}:  Assumptions,  Notation,  Implementation   --- Defines the general context for the tracking problems studied in later sections. Notation is  modified locally to accommodate the needs of each section.
    \end{itemize}
The next three sections treat the class of  probabilistic data association (PDA) filters. The    PDA filter corresponds to the classical single target problem and the  J-PDA filter to  the  multiple target  problem.   This class of filters estimate posterior probability density functions (pdfs). 

\begin{itemize}
    
    \item \underline{Section \ref{SSS}}:  Single Targets ---  uses a PGFL in the context of one target, one sensor, and one scan. Combinatorial problems arise in all but  the Bayes-Markov filter. 
    \item \underline{Section \ref{MSS}}: Labeled Multiple Targets  ---   multitarget versions of the single target problems of the previous section. Different  targets are uniquely identified by their labels.  
    \item \underline{Section \ref{MultipleSensors}}: Multisensor  Labeled    Multitarget ---   the labeled multiple target tracking problems of the previous section are discussed  for multiple sensor setting. 
    \end{itemize}
The next two sections discuss multiple target problems in which targets are unlabeled and their states can be superposed. They are called intensity filters (iFilters) because they  estimate  posterior intensity functions, i.e.,  the expected number of targets per unit state space. Single and multiple sensor problems are treated separarately.  

\begin{itemize}
\item \underline{Section \ref{iFilter}}:  Unlabeled  Multitarget ---   versions of the multitarget problems of the previous section in which the target states are treated as points in a common state space, i.e, targets are superposed.  Superposed targets are  unlabeled.   
\item{\underline{Section \ref{UnlabeledMultisensorMultitarget}}}: Multisensor Unlabeled  Multitarget --- the first subsection gives the multisensor versions of the problems  in Section VII. The second  discusses a  novel ``forensics'' problem in which the sensor labels of the measurements are removed in an effort to anonymize the sensors.  

\end{itemize}
The problems in the next several sections are either new  or have been treated by other methods in the literature.  

\begin{itemize}
{\color{black}    \item \underline{Section \ref{LabeledAndUnlabeled}}:} Hybrid  Labeled \&  Unlabeled Multitarget  --- tracking problems can combine both labeled  and  unlabeled targets. One problem tracks a single target-of-interest in a heterogeneous field of known and unknown targets, none of which are themselves tracked.  Another tracks a known number of targets that are observed not as points but as randomly  fluctuating sets of  highlights (e.g., specular reflections).    

{\color{black}\item \underline{Section \ref{MultipleScan}}:}    Labeled and Unlabeled Multitarget Trajectory --- A trajectory process is defined as a state-dependent branching process over a finite  interval.  A backward recursion for the PGFL of the trajectory process is derived.   

{\color{black}\item \underline{Section \ref{GFforAssignments}}:}  PGFs for   Assignment Probabilities ---  The  method is illustrated by several examples. The first two  derive  the PGF of the measurement-to-target assignment probabilities  for the  PDA  and  J-PDA filters. The other is new---it derives the PGF of the track-to-track assignment probabilities for  a labeled multisensor  problem.  

\end{itemize}
The next   section derives a principled approximation to filters from  their  PGFLs.  The method can be applied to all PGFL-based filters, including those known to have high computational complexity.   

\begin{itemize}
       {\color{black}\item \underline{Section \ref{saddlept}}:} Saddle Point Method  --- low computational complexity approximation is   derived for  NP-hard problem  of computing  particle weights for sequential importance resampling (SIR) particle filter implementations of  the J-PDA filter. 
\end{itemize}
The last section brings attention to the pair correlation function and  Palm processes, both of which can be formulated in terms of PGFLs and are relevant problems in tracking.  It is followed by an appendix that  reviews finite point processes, PGFs, and PGFLs.   
\begin{itemize}
{\color{black}\item \underline{Section \ref{conclude}}:} Concluding Remarks ---    surveys several combinatorially interesting statistics and  problems that fall outside  the scope of this paper. 
\end{itemize}

\begin{itemize}
    {\color{black}\item \underline{{Appendix}}}: ---  reviews    PGFs and PGFLs for finite point processes on  discrete and continuous spaces.   Bayes Theorem  for bivariate finite point processes is  stated in terms of the  PGF of the joint  process. 
\end{itemize}
The paper uses hyphenated acronyms to distinguish   the many different tracking filters   discussed in this paper.   The acronyms identify  the underlying  filter assumptions and are  consistent  with other widely used acronyms whenever possible.

\bigskip
{\color{black}
\section{Assumptions and Notation}
\label{AssocAndNotation}}

This papers uses  standard well-known  assumptions about targets and sensors.  They are enumerated here. The notation is general-purpose, but it is modified and extended as needed in subsequent sections.  All such modifications are  noted  explicitly in each section.       

\subsection{Target and Sensor Assumptions }
\label{Assumptions}
\begin{itemize}
    \item A target is characterized by a state $x$, where $x$ is a  point in a specified Euclidean state space $\mathcal X$. 
     
     \item Targets are uniquely identifiable if each  target has its own state space.  Such targets are  said to be labeled, a word that we use in the same sense that it is  used in the statistics literature.   Labeled targets can be heterogeneous, i.e., they can have different motion models state spaces $\mathcal X_n$, and measurement likelihood functions. The   spaces $\mathcal X_n$ can be   copies of the same space $\mathcal X$.
       
        \item Target    motion is Markovian, i.e., the pdf of the current  target state is conditioned  only on its previous state.   
        
        \item A target in state $x\in\mathcal X$ is  detected by a sensor if it generates {\it{at least}} one sensor measurement $y\in\mathcal Y$ in that sensor. The ``at most one measurement per target'' rule is a  special case.  The detection processes are independent from sensor to sensor.   
        \item If there are multiple sensors, their measurement scans are independent when conditioned on the targets that are present. 
        \item  The sensor locations (or, more generally, the sensor states) are known, unless stated otherwise. 
    \end{itemize}
\subsection{Measurement Assumptions}
\begin{itemize}
         \item Sensor measurements are points in a Euclidean space $\mathcal Y$. The space $\mathcal Y$  is assumed to be bounded.  Multiple sensors can be heterogeneous, i.e., they can have different spaces $\mathcal Y_\ell$. 
      \item A sensor  measurement scan comprises a finite set of sensor measurements that are made at the same moment in time. The set may be empty.   Different sensors are said to have different scans. Sensor scans are synchronous.   
      \item Every point in a  scan is produced by the sensor signal processor by one of two mechanisms: as a response to a target, or as a false alarm (FA) that is  unrelated to any target. 
    \item Sensor measurements  are not labeled, i.e., it is not known if a measurement is a false alarm or was generated by a target.  If there are multiple sensors, it is known which sensor generated which measurement, unless otherwise stated.
    \item False alarms  are modeled as finite point processes.   False alarm processes are independent  of all target measurement processes, and  are  independent  from sensor to sensor and from scan to scan. 
   
    \end{itemize}

\subsection{Notation}
\label{Notation}
\begin{itemize}
    \item For clarity, we reserve the index $n$ for target labels, $m$ for measurements,  $\ell$ for sensors, and $k$ for  the current scan.  The index $k$  is   included in the notation only when required for  clarity.  

     \item  A measurement scan is denoted by ${\mathbf y\subset \mathcal Y}$. If $\mathbf y$  has ${m\geqslant 1}$ measurements, we write ${\mathbf y=\{y_1,\ldots,y_m\}}$.  A sequence   of scans up to and including scan $k$  is denoted by  $\mathbb Y_k=(\mathbf y_1,\ldots,\mathbf y_k)$.  We define $\mathbb Y_0=\varnothing$.

     \item The (Bayes) posterior pdf of a  target  at scan $k$ conditioned on the measurement sequence $\mathbb Y_k$ is denoted by ${p_k^{}(x_k|\mathbb Y_{k})}$, ${x_k\in\mathcal X}$. We define ${p_{0}(x_{0}|\mathbb Y_{0})\equiv  \mu_0(x)}$. 
     
     \item Target motion is Markovian, so the predicted pdf of a   target is determined by the  posterior pdf at the previous scan:   \begin{equation}
    \label{SIR}
    \begin{aligned}
    p_k^{\!-}&(x_k|\mathbb Y_{k-1})\\
    &=\textstyle\int_{\mathcal X}p_{k}(x_k|x_{k-1}) p_{k-1}(x_{k-1}|\mathbb Y_{k-1}) \,\mathrm dx_{k-1},
    \end{aligned}
    \end{equation}
    where $p_{k}(x_k|x_{k-1})$ is  the Markov target transition function. 
The scan subscript $k$ is used primarily in Section \ref{MultipleScan}, so to simplify the  notation,  we  drop the subscript $k$ and the conditioning on $\mathbb Y_{k-1}$; thus, we write   ${p^{\!-}(x)\equiv p_k^{\!-}(x_k|\mathbb Y_{k-1})}$.  If there are ${N\geqslant 1}$ labeled  targets, we write ${p_n^-(x_n)}$, where ${x_n\in\mathcal X_n}$.  This usage should not cause confusion because,   by convention, we use $n$  for targets and $k$  for scans, and because in this paper we do not encounter a problem that requires us  to refer specifically to the pdf of target $n$ at time $k$.     
 
        \item $P\!d(x)$ is the probability of detecting a target at $x\in\mathcal X$. It is not a pdf, so  its integral over $\mathcal X$ need not equal  one. The non-detection probability is  ${Qd(x)=1-P\!d(x)}$. If  there are several labeled targets, we write  $P\!d_n(x_n)$ and $Qd_n(x_n)$, where $x_n\in\mathcal X_n$.  If  there is  more than one sensor,  we write   $P\!d_n^\ell(x_n)$ and  $Qd_n^\ell(x_n)$. 
    Finally, if  the targets are unlabeled, we cannot indicate target-dependent detection probabilities so we write $P\!d(x)$ and $Qd(x) $ if $L=1$ and  $P\!d^\ell(x)$ and $Qd^\ell(x) $ if $L>1$.

    \item $p(y|x)$ is the  likelihood of a sensor measurement ${y\in\mathcal Y}$ conditional on a target in state $x\in\mathcal X$.  If  there is more than one target, we write $p_n(y|x_n)$, where $x_n\in\mathcal X_n$. If  there is  more than one sensor,  we write  $p_n^\ell (y_\ell|x_n)$, where $y_\ell\in\mathcal Y_\ell$. It is a pdf since  $\int_{\mathcal Y_\ell}p_n^\ell(y_\ell|x_n)\,\mathrm dy_\ell = 1$ for all $x_n$. If  $\mathcal X_n\equiv \mathcal X$ for all $n$, then we  simplify the notation and write $p^\ell (y_\ell|x_n)$.

    \item  In IPDA filters, $\chi$ is the probability that a target is present, i.e., ${\chi=\Pr\{N=1\}}$. It is often called the target existence probability. 
   If  multiple targets are present, we write $\chi_n^{}$.
   \item The  arguments of PGFs and PGFLs are complex-valued  variables and functions, respectively.  In this paper we call them indeterminates,  in keeping with combinatorial    tradition.         \item  The function ${h:\mathcal X\rightarrow \mathbb C}$ denotes an indeterminate function for the PGFL of a target with state space $\mathcal X$.  If there are $N$ labeled targets and  target $n$ has state space $\mathcal X_n$,  the  indeterminate function for target $n$ is denoted by ${h_n:\mathcal X_n\rightarrow \mathbb C}$. The indeterminate functions of $N$ labeled targets are written  $h_{1:N}$.  
   \item Removing  the labels from $N$ labeled  targets gives $N$ unlabeled targets.  If the labeled target state spaces are identical, so that $\mathcal X_n=\mathcal X$ for all $n$, removing the labels   is equivalent to    making  the indeterminate functions identical, that is,  ${h_n=h}$ for all  $n$.        
    \item The function ${g:\mathcal Y\rightarrow \mathbb C}$ denotes an indeterminate function for a sensor with measurements  in the space  $\mathcal Y$.   If there are $L\geqslant 1$ possibly heterogeneous sensors, each with its own measurement space $\mathcal Y_\ell$, then $g_\ell:\mathcal Y_\ell\rightarrow \mathbb C$ is the  indeterminate function for sensor $\ell$.  A  sequence of  such functions is $g_{1:L}$.
    \item $\mathrm G^{{N}}(z)$ is the PGF of the number of targets, $N$, in a scan, where $z\in\mathbb C$ is an indeterminate variable. If $N$ is the number of targets in the previous scan, then $N^-$, is the ``predicted'' number at the current scan.    Its PGF is denoted $\mathrm G^{N^-}(z)$.
    
        \item $\mathrm G^{\mathrm{FA}}(w)$ is the PGF of the number of false alarms, where $w\in\mathbb C$ is an indeterminate variable. The number of false alarms can be arbitrarily distributed, but it is often taken to Poisson distributed, in which case $\lambda^{\mathrm{FA}}$ is the expected number  false alarms.  If  there is  more than one sensor,  we write $\lambda_\ell^{\mathrm{FA}}$.

\end{itemize}




\bigskip
{\color{black}
\section{Single Target Tracking}
\label{SSS}}

The Bayes-Markov (BM) filter is the foundational single target Bayesian  filtering problem  (see \cite{SSA} for a recent introductory discussion). The Bayes-Markov with Detection (BMD) filter adds the uncertainty of missed target detections to the BM problem.   Both assume that  false alarms are absent. The PDA filter  superposes the BMD target measurement process with an independent  false alarm process \cite{YBSTse1975}.  In this context, superposition means that  sensor measurements are unlabeled, i.e.,   it is not known which measurements are false alarms and which are not.  The PGFLs of these  filters are discussed in the first three subsections.   

The last two subsections add different kinds of uncertainty to the PDA problem.  The IPDA filter adds the uncertainty of target presence.  The interacting multiple model (IMM) technique   adds  uncertainty to the target motion model.  It can be added to the PDA and IPDA filters, but only the IMM-PDA filter is presented here.  

\subsection{Bayes-Markov (BM)}
\label{BayesMarkovBM}

It  assumes that  exactly one target is always present, that it always generates  one point measurement, and that there is precisely one measurement in the sensor scan. 
With these assumptions,  the  PGFL of the  measurement process, conditioned on $x$,  is given by 
\begin{equation}
\label{BasicMeasBM}
    \Psi^{\mathrm{BM}|x}(g)\equiv \textstyle\int_{\mathcal Y} g(y) p(y|x)\,\mathrm dy.
\end{equation} 
The PGFL  of the joint target-measurement process is  
\begin{equation}
 \label{BasicBM} 
\boxed{
\begin{aligned}
\Psi^{\mathrm{BM}}&\big(h,g\big)=\int_{\mathcal X} h(x) p^-(x) \Psi^{\mathrm{BM}|x}(g)\,\mathrm dx\\
 &=\int_{\mathcal X} h(x) p^-(x)\Big(\textstyle\int_{\mathcal Y} g(y) p(y|x)\,\mathrm dy\Big)\mathrm dx.
 \end{aligned}
 }
 \end{equation}
This PGFL applies to any number of targets and any number of measurements, however, {\it{the   only nonzero probability events are those with exactly one target and one measurement.}} (See the last paragraph in Section C of the Appendix.)  The PGFLs \eqref{BasicMeasBM}-\eqref{BasicBM} are  derived using the methods given in the Appendix (Sections B and C) via a gridded method in the limit as the size the grid cells goes to zero.   

\subsection{Bayes-Markov with Detection (BMD)} 

The BMD filter incorporates missed target detections into the  BM filter, that is, the  sensor may generate  {\it{at most}} one target measurement instead of exactly one.  Let $P\!d(x)$ denote the probability that a target in state $x\in\mathcal X$ is detected, i.e., that it  generates a measurement somewhere in $\mathcal Y$. The probability that it is not detected is  ${Qd(x)=1-Pd(x)}$.  This is a Bernoulli model of target detection where the probability of ``success'' is $P\!d(x)$. The Bernoulli PGF  (see Eqn. \eqref{BernoulliPGFAppdx} in the Appendix) generalizes  the PGFL of the conditional measurement process   from  \eqref{BasicMeasBM}  to     
\begin{align}
\label{BMDgf}
    \Psi^{\mathrm{BMD}|x}(g)& = 
    \mathrm G^{\mathrm {Bern}(P\!d(x))}\big(\textstyle\int_{\mathcal Y} g(y) p(y|x)\,\mathrm dy\big)\nonumber
    \\&=Qd(x)+P\!d(x)\textstyle\int_{\mathcal Y} g(y) p(y|x)\,\mathrm dy. 
\end{align}
It also generalizes  the joint PGFL  from \eqref{BasicBM} to 
\begin{equation}
\label{BasicBMD}
\boxed{\begin{aligned}
        &\Psi^{\mathrm{BMD}}(h,g) =  \int_{\mathcal X} h(x) p^-(x) \Psi^{\mathrm{BMD}|x}(g)\,\mathrm dx\\
        &\,\,\,\, =\int_{\mathcal X} \! h(x) p^-(x) \Big(Qd(x)+P\!d(x)\textstyle\int_{\mathcal Y} g(y) p(y|x)\,\mathrm dy\Big)\,\mathrm dx.
\end{aligned}
}
\end{equation}
The only nonzero probability events  are those with exactly one target and no more than one measurement.  

\subsection{Probabilistic Data Association (PDA)}
\label{PDAexample}
The classic  PDA filter \cite{YBSTse1975} is almost universally presented under the assumptions of linear-Gaussian target motion and measurement models and Poisson distributed false alarms. These assumptions are unnecessary and are not made in this paper.  

The PDA filter assumes that the sensor generates at most one measurement of the target, and that  measurement (if it is generated) is superposed with the sensor false alarm process.   False alarms are modeled as points in a realization of a (possibly nonhomogeneous)  finite point process on $\mathcal Y$. This means that (a) false alarms are i.i.d. points in $\mathcal Y$ with a pdf $p^{\mathrm{FA}}(y)$, (b)  false alarms are distinct with probability 1 because  $p^{\mathrm{FA}}(y)$ has no point masses,  and (c) the number $m$ of false alarms is a random non-negative integer with PGF 
\begin{align}
\label{generalFAprocess}
     \mathrm G^{\mathrm{FA}}(z)=\textstyle\sum_{m=0}^\infty \Pr\{\mathrm{Number\, of\, false\, alarms =\,}m\} \,z^m,
\end{align}
where $z$ is a complex-valued indeterminate variable. The PGFL of the false alarm process  is 
\begin{align}
\label{generalFA}
    \Psi^{\mathrm{FA}}(g)= \mathrm G^{\mathrm{FA}}\!\left(\textstyle\int_{\mathcal Y}g(y) p^{\mathrm{FA}}(y) \,\mathrm dy\right).
\end{align}
The false alarm  and  target measurement processes  are assumed independent, hence the PGFL of their superposition is the product,  
\begin{equation}
\label{PDA}
\boxed{
\begin{aligned}
    &\Psi^{\mathrm{PDA}}(h,g)=\Psi^{\mathrm{FA}}(g)\Psi^{\mathrm{BMD}}(h,g)\\
    &\quad=\mathrm G^{\mathrm{FA}}\!\left(\textstyle\int_{\mathcal Y}g(y) p^{\mathrm{FA}}(y) \,\mathrm dy\right)\\
    &\quad\,\,\,\,\,\,\,\times \!\int_{\!\mathcal X}\! h(x) p^{\!-}\!(x)\! \left(Qd(x)\!+\!P\!d(x)\!\textstyle\int_{\mathcal Y}  g(y) p(y|x)\,\mathrm dy\right)\mathrm dx.
\end{aligned}   
}
\end{equation}

The PGFL of the  Bayes posterior process,  conditioned on  the  measurements ${\mathbf y=\{y_1,\ldots,y_M\}\subset\mathcal Y}$,  is  the normalized mixed first-order  derivative of the joint PGFL with respect to $\mathbf y$. (The general method is outlined in the Appendix,   Eqns. \eqref{easyBi}-\eqref{BayesSec}.)  The measurement set $\bf y = \varnothing$ for ${M=0}$, so no derivatives are taken and the PGFL of the posterior process is evaluated at  ${g(\cdot)= 0}$, 
\begin{align}
    \label{PDAposterior0}
    \Psi^{\mathrm{PDA}}(h|\varnothing)&=\frac{ \Psi^{\mathrm{PDA}}(h,0)}{ \Psi^{\mathrm{PDA}}(1, 0)}=\frac{\int_{\mathcal X} h(x) p^{-}(x) Qd(x)\,\mathrm dx}{\int_{\mathcal X} p^{-}(x) Qd(x)\,\mathrm dx}.
\end{align}
For ${M\ge 1}$, we  find the secular function  by substituting  the weighted delta train (cf. \eqref{xxxx})
$$
g_{\mathrm{sec}}(y)=\textstyle\sum_{m=1}^M \gamma_m \delta_{y_m}(y)
$$ 
into  \eqref{PDA}, where ${\gamma=(\gamma_1,\ldots,\gamma_M)\in\mathbb C^M}$. The result is  denoted by 
\begin{equation}
\begin{aligned}
    \label{mumboPDA}
    \Psi_{\mathrm{sec}}^{\mathrm{PDA}}(h_{},\gamma)\equiv \mathrm G^{\mathrm{FA}}\left(\textstyle\sum_{m=1}^M a_m \gamma_m\right)\left(b_{0}+\textstyle\sum_{m=1}^M b_{m} \gamma_m\right),
\end{aligned}
\end{equation}
where, to simplify notation, we let 
\begin{subequations}
\label{cherry}
\begin{align}
  a_m&= p^{\mathrm{FA}}(y_{m}) \\
  b_{0}(h)&=\textstyle \int_{\mathcal X} h(x) \,p^{\!-}(x)\, Qd(x)\,\mathrm dx \\
    b_{m}(h)&=\textstyle \int_{\mathcal X} h(x) \,p^{\!-}(x) \,P\!d(x) \,p(y_{m}|x)\,\mathrm dx.    
\end{align}\end{subequations} 
A  straightforward calculation\footnote{
{If the function  ${f\!:\mathbb R\rightarrow \mathbb R}$ is $M$ times  differentiable at zero, then}

$\textstyle{\left.\frac{\partial^M}{\partial \gamma_1\cdots\partial \gamma_{M}}\right|_{\gamma_1=\cdots =\gamma_{M}=0}  f\!\left(\sum_{m=1}^M a_m \gamma_m\right)\!\! \left(b_0+\sum_{m=1}^M b_m \gamma_m\right) } \\
= f^{(M)}(0) \,b_0\textstyle{\prod_{m=1}^M a_m}+ \,\textstyle{f^{(M-1)}(0)\sum_{m=1}^M{\textstyle{\!\left(\prod_{m'=1,m'\ne m}^M a_{m'}\!\right)}}\,b_m}.$
} 
of the mixed first-order derivative  of $\eqref{mumboPDA}$ with respect to  $\gamma_{}$ evaluated at $\gamma = 0$ (cf. Eqn.  \eqref{doubel}) gives   
\begin{align}
\label{DofPDAsec}
    {\mathrm D}^{(0,{{\mathbf 1}_M)}} \Psi_{\mathrm{sec}}^{\mathrm{PDA}}(h,{\mathbf 0}_M)
     = C_0(h,\mathbf y) 
    \,+\sum_{m=1}^M C_m(h,\mathbf y)
\end{align}
where 
\begin{subequations}
\label{dullville}    \begin{align}
    C_0(h,\mathbf y)&\equiv M!\,\Pr\{M\}\, b_0(h)\,\textstyle  \prod_{m=1}^M a_m
    \\
C_m(h,\mathbf y)&\equiv (M\!-\!1)!\,\Pr\{M\!-\!1\} \,b_m(h)\,\textstyle\prod_{m'\ne m}^M
    a_{m'},
\end{align}
\end{subequations}
and   ${
\Pr\{m\} = \frac{1}{m!}\left.\frac{\mathrm d^m}{\mathrm d z^m}\right|_{z=0}\mathrm{G^{FA}}(z)}
$ is the probability of  ${m\geqslant 0}$ false alarms.  The  PGFL of the posterior process of the PDA filter is the normalized derivative. Assuming  ${p^{\mathrm{FA}}(y_{m}) > 0}$ for all $m$,  
\begin{align}
    \label{PDAposterior1}
    \Psi^{\mathrm{PDA}}(h|\mathbf{y}_{})&=\frac{{\mathrm D}^{(0,{\mathbf 1}_M)} \Psi_{\mathrm{sec}}^{\mathrm{PDA}}(h,{\mathbf 0}_M)}{{\mathrm D}^{(0,{\mathbf 1}_M)} \Psi_{\mathrm{sec}}^{\mathrm{PDA}}(1,{\mathbf 0}_M)}=\frac{\int_{\mathcal X} h(x) p^{\!-}(x)\mathcal L(\mathbf y|x)\,\mathrm dx}{\int_{\mathcal X} p^{\!-}(x)\mathcal L(\mathbf y|x)\,\mathrm dx},
\end{align}
where  the conditional measurement pdf is given  by 
\begin{align}
    \label{PDAlikely}
    \mathcal L(\mathbf y|x)=Qd(x)+\frac{\Pr\{M-1\}}{M\Pr\{M\}}\sum_{m=1}^M \frac{P\!d(x) p(y_m|x)}{p^{\mathrm FA}(y_m)}.
\end{align}
The nonzero probability events have  one target and up to ${1+F\!A_{\max}}$ measurements, where $F\!A_{\max}$ is the maximum number of false alarms. 

The posterior intensity of a target at $x\in\mathcal X$ is the normalized derivative of the secular function of \eqref{PDAposterior1}. There is exactly one target, by assumption, so the posterior intensity is identical to the posterior pdf.  Substituting  ${h_{\mathrm{sec}}(\cdot)=\alpha \delta_{x}(\cdot)}$ gives the secular function $\Psi^{\mathrm{PDA}}_{\mathrm{sec}}(\alpha|\mathbf y)$.   The   derivative with respect to $\alpha$  evaluated at $\alpha=0$ is the posterior pdf, 
\begin{align}
    \label{PDApdfExample}
p^{\mathrm{PDA}}(x|\mathbf{y}_{}) = \left.\frac{\mathrm d}{\mathrm d\alpha}\right|_{\alpha=0}\!\Psi^{\mathrm{PDA}}_{\mathrm{sec}}(\alpha|\mathbf y)=\frac{p^{\!-}(x)\mathcal L(\mathbf y|x)}{\int_{\mathcal X} p^{\!-}(x)\mathcal L(\mathbf y|x)\,\mathrm dx}.
\end{align}
The measurement to target assignment probabilities are computed from the PGFL by the method given in  Section \ref{labeledPDAassign}. 

The traditional false alarm process used in the PDA filter is the   homogeneous Poisson point process with the expected number of false alarms $\lambda^{\mathrm {FA}}$. Its PGFL is  
\begin{align}
\label{PPP}
    \Psi^{\mathrm{FA}}(g)=\mathrm G^{\mathrm{Pois}(\lambda^{\mathrm {FA}})}\!\left(\textstyle\int_{\mathcal Y}g(y) p^{\mathrm{FA}}(y) \,\mathrm dy\right) ,
\end{align}
where  $\mathrm G^{\mathrm{Pois}(\lambda^{\mathrm {FA}})}(\cdot)$ is the PGF given by \eqref{PoissonGF}.     For this process,  the constant in \eqref{PDAlikely} is ${\frac{\Pr\{M-1\}}{M\Pr\{M\}}=\frac{1}{\lambda^{\mathrm{FA}}}}$, assuming that $\lambda^{\mathrm {FA}}>0$.



\subsection{Integrated PDA  (IPDA)}
\label{IntegratedPDA(IPDA)}
The IPDA filter \cite{MES1994} extends the  modeling assumptions of the PDA filter by adding the hypothesis that the   target is either present or absent.  The probability that the target is present at the previous scan is denoted by $\chi$, and a two-state Markov chain on $\{0,1\}$ models  transitions between target  presence and absence.  (If the chain is not absorbing, the target can be present or absent intermittently over time.)  Let $\chi^{\!-}$ denote the predicted probability that the  number of targets is 1. By assumption there is at most one target, so   $1-\chi^{\!-}$ is the probability that the number of  targets is 0.  The IBMD (integrated BMD)  process includes the number of targets, which is either zero or one.  Its PGFL  is a slightly modified version of  \eqref{BasicBMD}:    
\begin{equation}
\label{BasicIBMD}
  \Psi^{\mathrm{IBMD}}(h,g)=1\!-\!\chi^{\!-}+\chi^{\!-}\int_{\mathcal X} h(x) p^{\!-}(x) \Psi^{\mathrm{BMD}|x}(g)\,\mathrm dx.
\end{equation}
Superposing this process with a false alarm process gives the PGFL of the IPDA filter \cite{MusSongStr} as
\begin{equation}
\label{IPDA}
\boxed{
\begin{aligned}
    &\Psi^{\mathrm{IPDA}}(h,g)=\Psi^{\mathrm{FA}}(g)\,\Psi^{\mathrm{IBMD}}(h,g)\\
    &\,\,\,\, =\Psi^{\mathrm{FA}}(g)\Big(1\!-\!\chi^{\!-}+\chi^{\!-}\int_{\mathcal X} h(x) p^{\!-}(x) \Psi^{\mathrm{BMD}|x}(g)\,\mathrm dx\Big).
\end{aligned}
}
\end{equation}
Because  this expression is equivalent to   
\begin{equation}
  \label{IPDAbernoulli}  \begin{aligned}
&\Psi^{\mathrm{IPDA}}(h,g) \\
    &\quad = \Psi^{\mathrm{FA}}(g)\,\mathrm G^{\mathrm {Bern}(\chi^{\!-})}\Big(\textstyle \int_{\!\mathcal X} h(x) p^{\!-}(x) \Psi^{\mathrm{BMD}|x}(g)\,\mathrm dx\Big), 
\end{aligned}
\end{equation}
the IPDA filter is sometimes called a Bernoulli filter.  

Different false alarm processes give  different 
IPDA filters. Substituting  \eqref{BMDgf} into \eqref{BasicIBMD} and using the PGFL \eqref{PPP} for Poisson false alarms gives the  PGFL of the traditional IPDA explicitly:  
\begin{align}
    \label{IPDAexplicit}
    &\Psi^{\mathrm{IPDA}}(h,g)=\exp\left(-\lambda^{\mathrm{FA}}+\lambda^{\mathrm{FA}}\textstyle\int_{\mathcal Y}\,g(y) p^{\mathrm{FA}}(y)\,\mathrm dy \right )\\
    &\!\times \!\!\left(\!1\!-\!\chi^{\!-}\!\!+\!\chi^{\!-}\!\int_{\!\mathcal X} \!h(x) p^{\!-}\!(x)\! \left(Qd(x)\!+\!P\!d(x)\!\textstyle\int_{\mathcal Y}  g(y) p(y|x)\,\mathrm dy\!\right)\!\mathrm dx\!\right)\!\!.\nonumber
\end{align}
Nonzero probability events for this PGFL comprise at most one target and any number of measurements. 

An  equivalent form of    \eqref{BasicIBMD} is    
\begin{align}
\label{NotSoBasicIBMD}
  \Psi^{\!\mathrm{IBMD}}&(h,g)=\int_{\mathcal X} p^{\!-}(x)\Big(1\!-\!\chi^{\!-}+\chi^{\!-}h(x)  \Psi^{\mathrm{BMD}|x}(g)\Big)\,\mathrm dx\nonumber 
  \\&=\int_{\mathcal X} p^{\!-}(x)\, \mathrm G^{\mathrm {Bern}(\chi^{\!-})} \Big(h(x)  \Psi^{\mathrm{BMD}|x}(g)\Big)\,\mathrm dx.
\end{align}
Using this  form in \eqref{IPDA} enables the PGFL  of the IPDA filter to accommodate  state-dependent existence probability functions $\chi=\chi(x)$; thus, \eqref{IPDAbernoulli} generalizes to 
\begin{equation}
   \label{IPDAlive}
   \begin{aligned}
    &\Psi^{\!\mathrm{IBMD}}(h,g)\\&=\Psi^{\mathrm{FA}}(g)\int_{\mathcal X} p^{\!-}(x)\, \mathrm G^{\mathrm {Bern}(\chi^{\!-}(x))} \Big(h(x)  \Psi^{\mathrm{BMD}|x}(g)\Big)\,\mathrm dx.
\end{aligned}
\end{equation}
A proof of the correctness of this legerdemain is given in \cite{2026Streit}.


\subsection{Interacting Multiple Model (IMM)}  
\label{IMM Filters}
IMMs are designed to  model  targets that may shift abruptly from one operating ``mode'' to another.  For example, a target  may maneuver by shifting    from constant velocity to  constant acceleration, or vice versa. Such shifts may occur at unknown times. The goal of the  IMM-PDA filter \cite{Blom},  \cite{BYBS} is  to track targets that are maneuvering, i.e., mode shifting, by incorporating an IMM into a PDA filter.   

The target IMM modes are labeled by the integers in the set  ${\mathcal S=\{1,\ldots,S\}}$, where the number of modes $S\geqslant 1$ is given. The state space of a  mode shifting target is  $\mathcal X\times \mathcal S$.  A discrete-continuous pdf on $\mathcal X\times \mathcal S$  has the form $p(x|s)\Pr\{s\}$, where  $p(x|s)$ is  a pdf on $\mathcal X$ for each $s\in \mathcal S$. The  integral  over $x$  and sum over $s$ is equal to one.  Mode shifts are modeled by a Markov chain on $\mathcal S$ with transition probability matrix  is $\Pi_k=\big[\pi_k^{is}\big]$, where 
\begin{align}
\label{PiYouCannotEatIt}
    \pi^{is}_k=\Pr\{\mathrm{mode}\!=\!s\,\mathrm{at\,time}\,{k}\,|\,\mathrm{mode}\!=\!i\,\mathrm{ at\,time}\,{k-1}\}.
\end{align}
Let  $w^i_{k-1}=\Pr\{i|\mathbb Y_{k-1}\}$ denote the conditional  probability that the target is in mode $i$ at time  $k-1$, and let   $w^{s-}_k$ denote the predicted probability that the target is in mode $s$ at time $k$. We  omit  the  dependence  on  $k$ from the notation and  write these probabilities as $w^i$ and $w^{s-}$. We also  write   $\Pi=\big[\pi^{is}\big]$.  From the Markov chain, we have     ${w^{s-}=\sum_{i=1}^S w^i \pi^{is}}$.


Modifying the target motion model   \eqref{SIR} to accommodate  mode $s$ gives  the predicted pdf of the target at the current scan as 
\begin{align}
    \label{SIRmode}
p_k^{s-}(x_k|\mathbb Y_{k-1})=\int_{\mathcal X}p_{k}^s(x_k|x_{k-1}) p_{k-1}^s(x_{k-1}|\mathbb Y_{k-1}) \,\mathrm dx_{k-1}.
\end{align}
We  omit  the  dependence  on  $k$ from the notation. The Bayesian recursion defines the  posterior  pdf at time ${k-1}$  to be    the prior pdf ${p^{s}(x)\equiv p_{k-1}^s(x_{k-1}|\mathbb Y_{k-1})}$ for time $k$. The predicted pdf is denoted $ p^{s-}(x)\equiv p_k^{s-}(x_k|\mathbb Y_{k-1})$. 

The prior pdf of the target in the IMM filter is  a list of mode-specific   pairs of probabilities and  pdfs, ${\{(w^s, p^s(x)):{1\leqslant s\leqslant S}\}}$. The  list is predicted to the current time, yielding the predicted list       ${\{(w^{s-},p^{s-}(x)):{1\leqslant s\leqslant S}\}}$.    

The target state space is ${\mathcal X\times \mathcal S}$, so the PGFL  uses the ordered pair of indeterminates   $(h(x),j_s)$, where $h(x)$ is the indeterminate function on $\mathcal X$ and $j_s$ is the  indeterminate complex-valued variable  for target mode $s$. As usual, $g(\cdot)$ is the indeterminate function for the measurement space $\mathcal Y$.   The PGFL of the IMM-PDA filter is 
\begin{equation}
    \label{IMMbasic}
\boxed{
\begin{aligned}
    \Psi^{\mathrm{IMM\mhyphen PDA}}&(h,j_{1:S},g)\!= \Psi^{\mathrm{FA}}(g) \,\sum_{s=1}^S \,j_s\,w^{s-} \!\!\int_{\!\mathcal X}\! h(x) p^{s-}(x) \\
    &\,\,\,\times\!\Big(Qd^s(x)+P\!d^s(x)\textstyle\int_{\mathcal Y} g(y) p^s(y|x)\,\mathrm dy\Big)\,\mathrm dx,
\end{aligned}
}
\end{equation}
where $p^s(y|x)$ is  the pdf of a measurement $y\in\mathcal Y$ conditioned on the  target in state $x$ and mode $s$, and $j_{1:S}=(j_1,\ldots,j_S)$. 

The   $j_s$  variables give insight into the filter. To see this, consider  the  case of no missed target detections, $P\!d(x)=1$, and no clutter, $\Psi^{\mathrm{FA}}(g)=1$.  In this case, there  is one measurement, say, $y\in\mathcal Y$. The PGFL of the state-mode process on   ${\mathcal X\times \mathcal S}$  conditioned on $y$ is (use Eqn.  \eqref{BayesIsBeautiful} in the Appendix)
\begin{align}
   \Psi^{\mathrm{Bayes}}\big(h,j_{1:S}
    \,\big|\,y\big)=\frac{\sum_{s=1}^S \,j_s\,w^{s-} \int_{\mathcal X} h(x) p^{s-}(x) p^s(y|x)\,\mathrm dx}{\sum_{s=1}^S \,w^{s-} \int_{\mathcal X}  p^{s-}(x) p^s(y|x)\,\mathrm dx}.\nonumber
\end{align}
The denominator    ${ \int_{\mathcal X}  p^{s-}(x) p^s(y|x)\,\mathrm dx\equiv \Lambda^s}$ is the  probability of $y$  conditioned on  target mode $s$.   Marginalizing over target state by setting $h(\cdot)=1$ gives the PGF of the mode process on $\mathcal S$ 
\begin{align}
    \Psi^{\mathrm{Bayes}}\big(j_{1:S}
    \,\big|\,y\big)=\sum_{s=1}^S\left(\frac{ \,w^{s-} \Lambda^s}{\sum_{\sigma=1}^S \,w^{\sigma-} \Lambda^\sigma}\right) j_s.\nonumber
\end{align}
The coefficient of $j_s$ is the Bayesian updated probability of mode $s$, that is,  $\Pr\{\mathrm{mode =}\,s\,|\,y\}$.  Similarly, marginalizing over the  modes by setting $j_s=1$ for all $s\in S$ gives the PGFL of the target process on $\mathcal X$  
\begin{align}
   \Psi^{\mathrm{Bayes}}\big(h
    \,\big|\,y\big)=\frac{\sum_{s=1}^S \,w^{s-} \int_{\mathcal X} h(x) p^{s-}(x) p^s(y|x)\,\mathrm dx}{\sum_{s=1}^S \,w^{s-} \int_{\mathcal X}  p^{s-}(x) p^s(y|x)\,\mathrm dx}.\nonumber
\end{align}
whose intensity function at $x\in\mathcal X$  is  (details omitted)  
\begin{equation}
\label{IMMtedium}
\begin{aligned}
p^{\mathrm{Bayes}}&\big(x
    \,\big|\,y\big)=\frac{ \sum_{s=1}^S\,w^{s-} p^{s-}(x) p^s(y|x)}{\sum_{\sigma=1}^S \,w^{\sigma-} \Lambda^\sigma}  \\
    &={\sum_{s=1}^S }\,\Pr\{\mathrm{mode =}\,s\,|\,y\}\,\frac{p^{s-}(x) p^s(y|x)}{\Lambda^s}.    
\end{aligned}
\end{equation}
The intensity is a pdf on $\mathcal X$ since, in this special case of no false alarms or missed detections,   there is exactly one target.  
The output of the IMM filter is a pdf on ${\mathcal X\times \mathcal S}$ defined by the list 
\begin{align*}
    {\left\{\left ( \Pr\{\mathrm{mode =}\,s\,|\,y\},\,\frac{p^{s-}(x) p^s(y|x)}{\Lambda^s}\right)\,:\,1\leqslant s\leqslant S\right\}}
\end{align*}
of Bayesian updated  mode probabilities and mode-specific filters.  


\bigskip
{\color{black}
\section{Labeled Multitarget Tracking}
\label{MSS}}

The single target filters in Section \ref{SSS}  generalize  to problems in which the number of targets, $N\geqslant 1$, is specified. The single target filter assumptions are  applied to each target. The J-PDA filter and the J-IPDA  filters include a false alarm model, while the joint BM and joint BMD filters do not.  In addition,  the  target measurement processes are assumed to be independent of each other, so that the  PGFL of the multitarget process is the product of single target  PGFLs.  The notation in this section is modified to accommodate multiple different target models.    

The  product form of the joint PGFL preserves the combinatorial character of the single target filters while also showcasing the  greater combinatorial complexity of their  multitarget counterparts.      

\subsection{Joint BM  (J-BM)}
The joint BM (J-BM) filter assumes that each target generates exactly one measurement.  Thus, it assigns $N$ measurements 1-to-1 to $N$ targets.  Let $\mathcal X_n$ denote the  state space for target $n$. All target measurements are in the space $\mathcal Y$.  The measurement likelihood functions are denoted $p_n(y|x_n)$, for  $y\in\mathcal Y$ and $x_n\in\mathcal X_n$.  Each target generates exactly one measurement, so the PGFL of the measurement process for target $n$ is, from \eqref{BasicMeasBM}, 
\begin{align}
   \label{BasicMeasBM(n)}
    \Psi^{\mathrm{BM}|x_n}_n(g)=\int_{\mathcal Y} g(y) p_n(y|x_n)\,\mathrm dy,\quad x_n\in\mathcal X_n,
\end{align}
and the PGFL  of the joint target-measurement process is  
\begin{equation}
\label{BasicBM(n)}
        \Psi^{\mathrm{BM}}_n(h_n,g)=\int_{\mathcal X_n} h_n(x_n) p_n^-(x_n) \Psi^{\mathrm{BM}|x_n}_n(g)\,\mathrm dx_n,
\end{equation}
where $h_n$ is the indeterminate function for target $n$.  The $N$ target-measurement processes are assumed  mutually independent, so the joint PGFL is the  product,   
\begin{align}
\label{GFBM1N}
\boxed{
\Psi^{\mathrm{J\mhyphen BM}}(h_{1:N},g)=\prod_{n=1}^N \Psi^{\mathrm{BM}}_n(h_n,g),
}
\end{align}
which reduces to the single target case \eqref{BasicBMD}  for $N=1$. Nonzero probability events for this PGFL   comprise exactly $N$ targets and $N$  measurements.  

\subsection{Joint BMD  (J-BMD)}
The joint BMD (J-BMD) filter differs from  the J-BM filter in that each target, independently of other targets,  may or may not generate a measurement.  Let $P\!d_n(x_n)$ denote the probability that a target in state $x_n\in\mathcal X_n$   generates a measurement somewhere in $\mathcal Y$, so that $1-P\!d_n(x_n)$ is the probability that it does not. Using the notation of \eqref{BasicMeasBM(n)} and \eqref{BasicBM(n)}, the PGFL of the conditional measurement process of the BMD filter for target $n$ is
\begin{subequations}
\label{bmdofn}
\begin{align}
    \Psi^{\mathrm{BMD}|x_n}_n&(g)=Qd_n(x_n)+P\!d_n(x_n)\Psi^{\mathrm{BM}|x_n}_n(g)\\
    &=Qd_n(x_n)+P\!d_n(x_n)\!\int_{\mathcal Y} g(y) p_n(y|x_n)\,\mathrm dy.
\end{align}
\end{subequations}
The joint PGFL of the BMD filter for target $n$ is  (cf. \eqref{BasicBMD})  
\begin{align}
\label{BasicBMDn}
\Psi^{\mathrm{BMD}}_n(h_n,g)=\int_{\mathcal X_n} h_n(x_n) p_n^-(x_n) \Psi^{\mathrm{BMD}|x_n}_n(g)\,\mathrm dx_n.
\end{align}
The processes are independent, by assumption, so the joint PGFL is 
\begin{equation}
\label{GFBM1Nmulti}
\boxed{
\Psi^{\mathrm{J\mhyphen BMD}}(h_{1:N},g)=\prod_{n=1}^N \Psi^{\mathrm{BMD}}_n(h_n,g).
}
\end{equation}
Nonzero probability events  have exactly $N$ targets and at most $N$  measurements.

\subsection{Joint PDA  (J-PDA)}
The J-PDA filter \cite{ybsTefSch} differs from the J-BMD  problem in only one way ---  the set of target measurements is superposed with a false alarm process. The false alarm and target processes are independent, so the PGFL is the product of \eqref{generalFA} and \eqref{GFBM1Nmulti}, 
\begin{equation}
\label{gfJPDA}
\boxed{
\begin{aligned}\Psi^{\mathrm{J\mhyphen PDA}}&(h_{1:N},g)=\Psi^{\mathrm{FA}}(g) \Psi^{\mathrm{J\mhyphen BMD}}(h_{1:N},g)\\
&=\Psi^{\mathrm{FA}}(g) \prod_{n=1}^N \Psi^{\mathrm{BMD}}_n(h_n,g).
\end{aligned}
}
\end{equation}
By inspection, it reduces to the PDA   for $N=1$.  Nonzero probability events for the PGFL \eqref{gfJPDA}  have exactly $N$ targets and, for Poisson false alarms, the traditional model, any number of  measurements.

Substituting $h_{n'}(\cdot)=1,\,n'\ne n,$ into \eqref{gfJPDA} gives the marginal PGFL for target $n$:
\begin{align}
    \label{margieJPDA}
\Psi^{\mathrm{J\mhyphen PDA}}&(h_{n},g)=\Psi^{\mathrm{FA}}(g) \Psi^{\mathrm{BMD}}_{n}(h_n,g)\prod_{\substack{n'\ne n}} \Psi^{\mathrm{BMD}}_{n'}(1,g).
\end{align}
The PGFL of the posterior of the marginal  process is  $\Psi^{\mathrm{J\mhyphen PDA}}(h_{n} |\mathbf{y})$, where  $\mathbf{y}$ is the  scan measurement set (cf. \eqref{BayesIsBeautiful}).  
To close the Bayes recursion, the J-PDA filter approximates the posterior process by the product of its $N$ marginal processes.  In PGFL terms, this gives   
\begin{align}
    \label{meanfieldjpda}
    \Psi^{\mathrm{J\mhyphen PDA}}(h_{1:N}|\mathbf{y})\approx \textstyle \prod_{n=1}^N\Psi^{\mathrm{J\mhyphen PDA}}(h_n|\mathbf{y}).
\end{align}
Approximations of this kind are known in the statistical literature as  ``mean field'' approximations.  The approximation, in effect, reimposes the target independence assumption  on the joint posterior pdf.  

The  J-PDA filter is impractical for large $N$ because of its high computational complexity--it is NP-hard. To see why it is so difficult, it suffices to study the J-PDA measurement to target assignment problem.  Details are given in  Section  \ref{labeledJPDA}, but said simply, the PGFL \eqref{gfJPDA} leads to  expressions for the assignment probabilities that, even for no false alarms and no missed detections, are equivalent to computing  a matrix permanent and/or  its generalization. The matrix permanent is an exponentially hard calculation.   

\subsection{Joint IPDA  (J-IPDA)}
\label{JIPDA}
The J-IPDA model  \cite{2004MusickiEvans} differs from  J-IBMD model in that the set of target measurements is superposed with an independent false alarm process.  Using the general false alarm process \eqref{generalFA},  the PGFL of the predicted  J-IPDA multitarget-measurement process   is 
\begin{align}
\label{gfJIPDA}
\boxed{
\Psi^{\mathrm{J\mhyphen IPDA}}(h_{1:N},g)=\Psi^{\mathrm{FA}}(g) \prod_{n=1}^N \Psi^{\mathrm{IBMD}}_n(h_n,g)\,,
}
\end{align}
where, using \eqref{BasicBMDn} and adding a subscript $n$ to  \eqref{IPDA}-\eqref{IPDAbernoulli},   
\begin{equation}
  \label{BasicIBMDn}  
\begin{aligned}
&\Psi^{\mathrm{IBMD}}_n(h_n,g)\\
&\qquad = 1\!-\!\chi_n^{\!-}+\chi_n^{\!-}\int_{\mathcal X_n} h_n(x) p_n^{\!-}(x) \Psi_n^{\mathrm{BMD}|x}(g)\,\mathrm dx\\\
&\qquad = \mathrm G^{\mathrm{Bern}(\chi_n^{\!-})}\Big(\textstyle \int_{\mathcal X_n} h_n(x) p_n^{\!-}(x) \Psi^{\mathrm{BMD}|x}_n(g)\,\mathrm dx\Big).
\end{aligned}
\end{equation}
The product of the $N$ labeled target processes in \eqref{gfJIPDA} is called a  multiBernoulli process.  For a Poisson point process  model of false alarms \eqref{PPP}, the nonzero probability events  have at most $N$ targets and any number of  measurements. 

The  PGFL of the posterior J-IPDA process is the normalized   derivative (cf. Eqn. \eqref{BayesIsBeautiful}) of \eqref{gfJIPDA}. For   Poisson false alarms, the posterior process is  a labeled multiBernoulli mixture (LMBM).  To understand what this acronym means, we calculate the derivative of \eqref{gfJIPDA} for  the measurement set  ${\mathbf y=\{y_1,\ldots,y_M\}\neq \varnothing}$. Substituting the weighted delta train (cf. \eqref{xxxx})
$$
g_{\mathrm{sec}}(y)=\textstyle\sum_{m=1}^M \gamma_m \delta_{y_m}(y)
$$ 
into  \eqref{gfJIPDA}, where ${\gamma=(\gamma_1,\ldots,\gamma_M)\in\mathbb C^M}$, gives the secular function   
\begin{equation}
\begin{aligned}
    \label{mumbo}
    &\Psi_{\mathrm{sec}}^{\mathrm{J\mhyphen IPDA}}(h_{1:N},\gamma)\\
    &\qquad \equiv \exp\Big(a_0+\sum_{m=1}^M a_m \gamma_m\Big)\prod_{n=1}^N\Big(b_{n}+\sum_{m=1}^M b_{mn} \gamma_m\Big),
\end{aligned}
\end{equation}
where, to simplify notation, we let 
\begin{align*}
  a_0 &=-\lambda^{\mathrm{FA}}\\
  a_m&= \lambda^{\mathrm{FA}} p^{\mathrm{FA}}(y_{m}) \\
  b_{n}&=1-\chi_n^-+\chi_n^-\textstyle \int_{\mathcal X_n}h_n(x) p_n^{\!-}(x)\, Qd_n(x)\,\mathrm dx \\
    b_{mn}&=\chi_n^-\textstyle \int_{\mathcal X_n} h_n(x) p_n^-(x) \,P\!d_n(x) \,p_n(y_{m}|x)\,\mathrm dx.    
\end{align*}
Let  $\kappa_{n}=\textstyle \int_{\mathcal X_n} p_n^{\!-}(x)\, Qd_n(x)\,\mathrm dx$ and 
$$
{\nu_{n}= \frac{ \kappa_{n}\chi_n^-}{1-\chi_n^-+\kappa_{n}\chi_n^-}}.
$$
The   $b_{n}$ terms are proportional to the PGFL of a Bernoulli process,   
$$
b_{n}=\big(1-\chi_n^-+\kappa_{n} \chi_n^- \big) \Psi_n^{\mathrm{Bern}(\nu_{n})}\!\left(\frac{ \int_{\mathcal X_n} h_n(x) p_n^{\!-}(x)\, Qd_n(x)\,\mathrm dx}{ \int_{\mathcal X_n} p_n^{\!-}(x)\, Qd_n(x)\,\mathrm dx}\right)\!.
$$
The  $b_{mn}$ terms are also proportional to  Bernoulli processes, 
$$
b_{mn}=\kappa_{mn}  \chi_n^- \,\Psi_n^{\mathrm{Bern}(1)}\!\left(\!\frac{ \int_{\mathcal X_n} h_n(x)  p_n^-(x) \,P\!d_n(x) \,p_n(y_{m}|x)\,\mathrm dx}{ \int_{\mathcal X_n}  p_n^-(x) \,P\!d_n(x) \,p_n(y_{m}|x)\,\mathrm dx}\right)\!,
$$
where  $\kappa_{mn}=\textstyle  \int_{\mathcal X_n}  p_n^-(x) \,P\!d_n(x) \,p_n(y_{m}|x)\,\mathrm dx$. The mixed first-order derivative of \eqref{mumbo} at $\gamma={\mathbf 0}_M=(0,\ldots,0)\in\mathbb C^M$ is calculated  using the Leibniz rule.\footnote{Let ${f(x)=f_0(x) f_1(x)\cdots f_N(x)}$, $x\in\mathbb C^M$.   The Leibniz rule for the general mixed derivative of $f$ of order $\alpha\in\mathbb N^M$ is, using multi-index notation,  $f^\alpha=\sum_{\beta_1+\cdots+\beta_N\leqslant\, \alpha}\binom{\alpha}{\beta_1\cdots \beta_N} \, f_0^{\alpha-\beta_1\cdots -\beta_N} f_1^{\beta_1}\cdots f_N^{\beta_N}$.} 

To state the derivative, we first define $\Theta$ to be the set of all ${\{0,1\}}$-matrices of size  ${M\times N}$ each of  whose columns and rows  sum to at most 1. The set $\Theta$ is partitioned into subsets of matrices denoted by   ${\Theta(t),\, t\geqslant 0}$.  A matrix $\theta$ is in  ${\Theta(t)\subset \Theta}$ if and only if $\theta$ has exactly $t$ columns that sum to 1.  The set of indices of the nonzero  columns of  ${\theta\in\Theta(t)}$  is denoted ${\mathcal I\big(\theta\in\Theta(t)\big)}$, and  the set  ${\mathcal J\big(\theta\in\Theta(t)\big)}$ denotes the  other ${N-t}$ indices. For ${k=0}$, we define ${\mathcal I\big(\theta\in\Theta(0)\big)=\varnothing}$  and  ${\mathcal J\big(\theta\in\Theta(0)\big)=\{1,\ldots,N\}}$. Finally, the row in which the nonzero entry of  column  $t\in \mathcal I(\theta)$  occurs is denoted  $m_{\theta}(t)$.    With this notation, a tedious calculation (see  \cite[Appx. C.4]{2021ACBook} for details)   using the Leibniz rule shows that  the derivative of \eqref{mumbo} at $\gamma={\mathbf 0}_M$ is
\begin{equation}
  \label{mixture}  
\begin{aligned}
    &\mathrm D^{({\mathbf 0}_N,{\mathbf 1}_M)}\Psi_{\mathrm{sec}}^{\mathrm{J\mhyphen IPDA}}(h_{1:N},{\mathbf 0}_M) \\
    &\quad =\kappa \sum_{t=0}^{\min\{M,N\}} \sum_{\theta\in\Theta(t)}\left( \prod_{s\in\mathcal J(\theta)} b_{s}\right) \left( \prod_{s\in\mathcal I(\theta)} \frac{b_{m_{\theta}(s),t}}{a_{m_{\theta}(s)}} \right), 
\end{aligned}
\end{equation}
where $\kappa=e^{a_0} a_1\cdots a_M$.  Normalizing this derivative by dividing by its value at  ${h_n(x_n)\equiv 1,\, n=1,\ldots, N,}$ gives   the PGFL of the posterior J-IPDA process,  $\Psi^{\mathrm{J\mhyphen IPDA}}(h_{1:N}|\,\mathbf y_{})$.  The $b_{n}$ and $b_{mn}$ terms are, as noted above, proportional to  PGFLs of Bernoulli processes, so the posterior PGFL is the weighted sum of products of  PGFLs of mutually independent   Bernoulli processes. The superposition of independent Bernoulli processes is called a labeled   multiBernoulli (LMB) process. In this terminology, the  exact posterior PGFL is a probabilistic mixture of LMB processes, otherwise known as an   LMB mixture (LMBM) process. 

The order of an LMB process is defined to be the number of Bernoulli processes in the LMB.  The LMBs in the sum over $\Theta(t)$ are all of order $t$, so the derivative  \eqref{mixture} is an expansion of LMB processes of increasing order.  

If the PGFL of the prior process is  a LMBM process  and the the false alarm process is Poisson,  the posterior process is  again an  LMBM   process.  Each multiBernoulli process in the LMBM prior contributes a LMBM process to the final sum, so the posterior PGFL is a much larger LMBM process than the prior process.  The Bayesian filter recursion is, therefore, {\it{not}} closed in the strict statistical sense  because the number of terms in the posterior LMBM process grows rapidly  over time. 
(Note:  The  same issue appears in classical Gaussian sum filters.) 

To close the Bayes recursion it is necessary reduce the number of terms in the exact posterior LMBM process to match the order of the prior  LMB process.  One way to do this is discussed in \cite{2014ReuterVoVoDietmeyer}, where  it is called  the  LMB filter.  The PGFLs of this precess is identical to the PGFL of the J-IPDA filter.  

\subsection{Joint Interacting Multiple Model}
\label{JointIMMfilters}
The IMM  version of the  J-PDA filter allows different targets to have different modes, i.e., targets can maneuver in different ways. The modes for target $n$ are labeled $\mathcal S_n=\{\sigma_n^1,\ldots,\sigma_n^{S_n}\},\,S_n\geqslant 1$.  Mode switching for target $n$ is modeled by a target-specific Markov chain on $\mathcal S_n$ with  transition probability matrix $\Pi_n=\big[\pi_n^{ij}\big]$, where the definition of $\pi_n^{ij}\equiv \pi_n^{ij}(k)$ parallels that of  \eqref{PiYouCannotEatIt}.   The predicted  probability $w_n^{s-}$ that  target $n$ is in mode $\sigma_n^s\in\mathcal S_n$ at the current scan is, by the Markov chain,    ${w_n^{s-}=\sum_{j=1}^{S_n} w_n^j \pi_n^{js}}$, where $w_n^j$ is the probability that target $n$ is in mode $\sigma_n^j\in\mathcal S_n$ at the previous scan.

The  predicted mode-specific target pdfs  have the same form as in  single target IMM-PDA \eqref{SIRmode}, although the target motion and measurement models can in general be different for different targets. Simplifying  notation, as was done for single target IMM, we denote the pdf of target $n$ in mode $\sigma_n^s$ by  $p_n^{s-}(x)$, $1\leqslant s\leqslant S_n$.   

To write the PGFL, we use an  indeterminate function $h_n(\cdot)$ for target $n$ and an  indeterminate complex-valued variable ${{j}_n^s}$ to label  mode $s$ of target $n$. Let $J=(j_1,\ldots,j_N\}$, where $j_n\equiv(j_n^1,\ldots,j_n^{S_n})$. The PGFL of the IMM-J-PDA filter is 
\begin{equation}
    \label{IMMJPDAbasic}
\boxed{
\begin{aligned}
    &\Psi^{\mathrm{IMM\mhyphen J\mhyphen PDA}}(h_{1:N},J,g)\!= \!\Psi^{\mathrm{FA}}(g)\!\prod_{n=1}^N\!\!\left\{\sum_{s=1}^{S_n} j_n^s\,w_n^{s-} \!\!\!\int_{\!\mathcal X}\!\! h_n(x) \right. \\
    &\quad\Bigg. \times\, p_n^{s-}(x) \Big(Qd_n(x)\!+\!P\!d_n(x)\!\textstyle\int_{\mathcal Y} g(y) p_n^s(y|x)\,\mathrm dy\Big)\,\mathrm dx\Bigg\},
\end{aligned}
}
\end{equation}
where $p_n^s(y|x)$ is  the pdf of a measurement $y\in\mathcal Y$ conditioned on  target $n$ in state $x$ and mode $s$.

To find  the marginal PGFL for the IMM filter for target $n$ with modes in $\mathcal S_n$,  substitute $h_{n'}(\cdot)=1$ and $j_{n'}^s=1$ for  $n'\ne n$ into \eqref{IMMJPDAbasic}.  The resulting PGFL,  $\Psi^{\mathrm{IMM\mhyphen J\mhyphen PDA}}(h_{n},j_n,g)$, depends only on the function $h_n$ and the vector of mode indices $j_n$. It is the IMM extension of the J-PDA result \eqref{margieJPDA}. The mean field approximation, 
\begin{align}
    \Psi^{\mathrm{IMM\mhyphen J\mhyphen PDA}}(h_{1:N},J,g) \approx \prod_{n=1}^N \Psi^{\mathrm{IMM\mhyphen J\mhyphen PDA}}(h_{n},j_n,g),
\end{align}
closes the Bayes recursion.  

\bigskip
{\color{black}
\section{Multisensor Labeled  Multitarget Tracking}
\label{MultipleSensors}}

The sensor origin of every measurement is assumed known.  The sensor locations (or, more generally, the sensor states) are known.  The number of sensors is $L\geqslant 1$. The measurement space of sensor $\ell$ is denoted    $\mathcal Y_\ell$.   The state space of target $n$ is $\mathcal X_n$.  Denote the probability of  sensor $\ell$ detecting  target $n$ in state $x_n\in\mathcal X_n$  by  $P\!d_n^\ell(x_n)$, and let $Qd_n^\ell(x_n)=1-P\!d_n^\ell(x_n)$.  Denote the likelihood function of sensor $\ell$ by $p_n^\ell(y|x_n),\,y\in\mathcal Y_\ell$.  Sensor measurement sets are assumed to be  independent of each other when conditioned on the set of target states. 

The sensor measurement sets are  assumed to be communicated to  a ``fusion center'' so that they can be processed jointly. The PGFLs discussed in this section are those of the fusion center.  It is assumed that the sensors are  registered, that is, they are mapped into a common frame of reference to eliminate systematic nonrandom biases in the measurements.

\subsection{Multisensor  J-BMD} 
\label{MultiSensorBMD}
The MS-J-BMD (multisensor J-BMD) filter  subsumes  the MS-BM (multisensor BM) and MS-BMD  (multisensor BMD) filters as special cases. MS-J-BMD  superposes $N\geqslant 1$ independent multisensor BMD processes.   The targets are  labeled.  We  assume there are no false alarms.  Extending the  single sensor BMD  notation \eqref{bmdofn} gives  the  PGFL of the conditional measurement process of the BMD for target $n$ in  sensor $\ell$ as  
\begin{align}
    \label{bmdofnSensorEll}
    \Psi^{\mathrm{BMD}|x_n}_{\ell|n}(g_\ell)=Qd_n^\ell(x_n)+P\!d_n^\ell(x_n)\!\!\int_{\mathcal Y_\ell} g_\ell(y) p_n^\ell(y|x_n)\,\mathrm dy.
\end{align}
The measurement sets are independent conditional on $x_n$, so the PGFL for  target $n$  in each of the  $L$ sensors is (cf. \eqref{BasicBMDn})  
\begin{align}
\Psi^{\mathrm{BMD}}_{1:L|n}(h_n,g_{1:L})= \int_{\mathcal X_n} \!\!h_n(x_n) p_n^{\!-}(x_n)\prod_{\ell=1}^L \Psi^{\mathrm{BMD}|x_n}_{\ell|n}(g_\ell)\,\mathrm dx_n,\nonumber
\end{align}
where $p_n^{\!-}(x_n)$ is the prior pdf for target $n$.  A nonzero probability event for this conditional PGFL    comprises exactly one target and at most one measurement in each on the $L$ sensors. Superposing these multisensor BMD processes defines the  multisensor problem. By independence, the  joint PGFL of MS-J-BMD is 
\begin{equation}
\label{GFBM1NmultiEll}
\boxed{
\begin{aligned}
    \Psi&^{\mathrm{MS\mhyphen J\mhyphen BMD}}(h_{1:N},g_{1:L})=\prod_{n=1}^N  \Psi^{\mathrm{BMD}}_{1:L|n}(h_{n},g_{1:L})\\
&\quad = \prod_{n=1}^N \int_{\mathcal X_n} \! h_n(x_n) p_n^-(x_n) \prod_{\ell=1}^L\Psi^{\mathrm{BMD}|x_n}_{\ell|n}(g_\ell)\,\mathrm dx_n.
\end{aligned}
}
\end{equation}
Nonzero probability events   for MS-J-BMD have exactly $N$ targets and at most $N$  measurements in each of the $L$ sensors.  

\subsection{Multisensor  J-PDA,  J-IPDA, IMM-J-PDA}
\label{MultiSensorJPDA}
The PGFL of the  MS-J-PDA (multisensor J-PDA) filter is obtained by  superposing MS-J-BMD with an independent false alarm process in each sensor.   The false alarm processes may vary from sensor to sensor.  The PGF of the number $m$ of false alarms in  sensor $\ell$ (cf. \eqref{generalFAprocess}) is denoted by 
\begin{align}
\label{generalFAprocessEll}
     \mathrm G^{\mathrm{FA}}_\ell(z_\ell)=\textstyle\sum_{m=0}^\infty \Pr\{m\,|\,\mathrm{sensor}\,\, \ell\}\, z^m_\ell,
\end{align} 
where $z_\ell$ is the indeterminate variable.  The PGFL of the false alarm point process for sensor $\ell$ is (cf. \eqref{generalFA})
\begin{align}
\label{generalFAEll}
    \Psi^{\mathrm{FA}}_\ell(g_\ell)= \mathrm G^{\mathrm{FA}}_\ell\!\left(\textstyle\int_{\mathcal Y_\ell} g_\ell(y) p^{\mathrm{FA}}_\ell(y) \,\mathrm dy\right).
\end{align}
These processes are independent, so the joint PGFL is 
\begin{align}
\label{JointFalseAla}
    \Psi^{\mathrm{FA}}_{1:L}(g_{1:L})=\prod_{\ell=1}^L \Psi^{\mathrm{FA}}_\ell(g_\ell).
\end{align}
The false alarm processes and the target measurement processes are assumed independent, so the PGF of MS-J-PDA is (cf. \eqref{gfJPDA}) 
\begin{equation}
\label{gfJPDAELL}
\boxed{
\begin{aligned}
\Psi^{\mathrm{MS\mhyphen J\mhyphen PDA}}&(h_{1:N},g_{1:L})\\
&=\Psi^{\mathrm{FA}}_{}(g_{1:L})  \Psi^{\mathrm{MS\mhyphen J\mhyphen BMD}}(h_{1:N},g_{1:L}).
\end{aligned}
}
\end{equation}
When all sensors have Poisson false alarms, the nonzero probability events for MS-J-PDA comprise exactly $N$ targets and any number of  measurements in each sensor.

The  multisensor version of J-IPDA (MS-J-IPDA)  has exactly the same form, except that  $\Psi^{\mathrm{BMD}}_{\ell|n}(h_n,g_\ell)$ is replaced by the appropriate multisensor IBMD version of  \eqref{BasicIBMDn}, the PGFL for measurements of target $n$ in sensor $\ell$.  With Poisson false alarms,  the nonzero probability events for MS-J-IPDA  have at most $N$ targets and any number of  measurements in each sensor.  

Analogous comments hold for the IMM-J-PDA. Paralleling the discussion above leads  to the multisensor version of IMM-J-PDA, which has the unwieldy acronym MS-IMM-J-PDA.    

\bigskip
{\color{black}
\section{Unlabeled Multitarget Tracking}
\label{iFilter}
}

This section discusses tracking problems in which target  labels are unavailable.  
The absence  of  labels fundamentally  changes the phenomenology of a Bayesian  tracking problem.  An ordered {\it{list}} of labeled targets when stripped of the labels 
is equivalent to an unordered {\it{set}} of indistinguishable points. Likewise, without measurement labels, which measurements  
are false alarms and which  (if any) correspond to which target is unknown.   

The target and measurement processes in unlabeled tracking problems are   random variables called finite point processes.   
It is shown using Bayes Theorem (see Eqn. \eqref{BayesIsBeautiful}  of the Appendix) that the   posterior target process is a finite point process whose PGFL is a normalized mixed derivative of the  PGFL of the joint target-measurement process. 

Summary statistics of the posterior process are computed from the posterior PGFL and used in practical applications.  The most commonly  used statistic is the intensity function, i.e., the expected number of targets per unit state space conditioned on the measurement set.  An  intensity filter, or iFilter,  computes  the   intensity function of the posterior  point process.   Other  summary statistics are also useful but are outside the scope of this paper; they  are mentioned in the concluding remarks of Section \ref{conclude}.  

The PGFL for  unlabeled target problems is derived from the PGFL of the labeled versions   merely  by replacing the  functions $h_n$ in the labeled PGFL with the same function $h$. Using the same function $h$ for all targets is equivalent to superposing realizations of the target processes (see Appendix).  The  PGFL of the unlabeled, or superposed, problem  has fewer variables and  simpler derivatives.  This often---but not always---leads to filters with lower computational complexity than  their labeled counterparts.

\bigskip
\subsection{Unlabeled J-BM Targets with No False Alarms}
\label{SuperBM} 
The PGFL for the labeled J-BM filter is given by \eqref{GFBM1N}.  Setting ${h_n=h}$  gives  the PGFL of the U-J-BM (unlabeled J-BM)  filter,  
\begin{align}
\label{superGFBM1N}
\widetilde{\Psi}^{\mathrm{U\mhyphen J\mhyphen BM}}(h_{},g)=\prod_{n=1}^N \Psi^{\mathrm{BM}}_n(h,g),
\end{align}
where the PGFL \eqref{BasicBM(n)} is unchanged except that the indeterminate function is now $h$. To be clear, the BM  models may differ from target to target but which target generates which measurement is unknown.  

The complexity of the problem is significantly reduced when we  take the BM models to be identical; i.e., we  assume that $\mathcal X_n=\mathcal X$, $p_n^-(x_n)=p^{-}(x)$, $P\!d_n(x)=P\!d(x)$, and  $p_n(y|x)=p(y|x)$  for all $n$. With these assumptions it  follows that  $\Psi^{\mathrm{BM}}_n(h,g)$ is identical to $\Psi^{\mathrm{BM}}(h,g)$ as defined in \eqref{BasicBM}.  Using this and \eqref{superGFBM1N},  the PGFL of the  superposed U-J-BM as becomes
\begin{align}
\label{SuperGFBM1N}
\boxed{
\Psi^{\mathrm{\mathrm{U\mhyphen J\mhyphen BM}}}(h_{},g)=\big( \Psi^{\mathrm{BM}}(h,g)\big)^{\!N}.
}
\end{align}
The only nonzero probability events have exactly $N$ targets and $N$ measurements. The difference between J-BM and U-J-BM   is that the points in U-J-BM are not identifiable, i.e., which target generated which measurement is unknown.   

Assume now that the number of BM targets is random, and let $N$ be the number of targets at the previous scan. The number of targets predicted to the current scan is denoted by $N^-$.  Denote the PGFs by  $\mathrm G^N(z)$ and $\mathrm G^{N^-}(z)$, respectively, where $z$ is the complex-valued indeterminate variable.  The   PGF of $N^-$,   
\begin{align}
\label{cards}
     \mathrm G^{N^-}\!(z) = \textstyle \sum_{n=0}^\infty \Pr\{N^-\!=n\}\,z^n,
\end{align}
is arbitrary here, but in practice it is derived from the distribution of $N$ in a manner suited to the application.  The   U-J-BM filter with a random number of targets is denoted RU-J-BM.  Its PGFL is 
\begin{align}
    \Psi^{\mathrm{RU\mhyphen J\mhyphen BM}}(h,g) = \mathrm G^{{N}^-}\big(\Psi^{\mathrm{BM}}(h,g)\big).
\end{align}
For a Poisson number of targets,
\begin{equation}\label{okthisone}
\begin{split}
    &\Psi^{\mathrm{RU\mhyphen J\mhyphen BM}}(h,g) \\
    &\qquad =\exp\left(-\lambda^{N^-}\!+\,\lambda^{N^-}\!\int_{\mathcal X} h(x) p^{\!-}(x) \Psi^{\mathrm{BM}|x}(g)\,\mathrm dx \right),
\end{split}
\end{equation}
where $\lambda^{N^-}$ is the predicted expected number of targets,  and  $p^{\!-}(x)$ is the predicted pdf of the i.i.d.  sample points of the target PPP.  

The nonzero probability events  have a random number of unlabeled targets and a matching number of measurements. To see this analytically, note  that the intensity function of the Bayes posterior process conditioned on the measurements $\mathbf y =\{y_1,\ldots,y_M\}$ is 
\begin{align}
\label{simpleSimon}
\Lambda^{\mathrm {Bayes}}(x|\mathbf  y)={\sum_{m=1}^M} \frac{p^{-}(x) p(y_m|x)}{\int_{\mathcal X}p^{-}(x)p(y_m|x)\,\mathrm dx}.
\end{align}
The derivation uses \eqref{BayesIsBeautiful} and \eqref{PGFL12} of the appendix and is omitted.  

\subsection{Unlabeled J-BMD Targets with No False Alarms} 
\label{multiBernoulli}
The  number  of targets, $N$, is assumed to be fixed and unchanging, hence the predicted PGF is  ${\mathrm  G^{{N}^-}(z)=z^N}$.  We superpose the targets and make the same assumptions as the U-BM filter, adding the condition that $P\!d_n(x)=P\!d(x)$, so  we can use the BMD model \eqref{BasicBMD} instead of \eqref{BasicBM}. The BMD models are the same for all targets, so PGFL for the U-J-BMD (unlabeled J-BMD) is   
\begin{align}
 \label{SuperposedJPDAtheRealOne}
 \boxed{    \Psi^{\mathrm{U\mhyphen J\mhyphen BMD}}(h_{},g)=\big( \Psi^{\mathrm{BMD}}(h,g)\big)^N,
 }
 \end{align}
where $\Psi^{\mathrm{BMD}}(h,g)$ is given by \eqref{BasicBMD}.The  nonzero probability events using this PGFL have exactly $N$ targets and at most $N$ measurements. The input to the filter is the  pdf of the predicted point process, $p^-(x)$.  

Suppose now that $N$ is random.  If   the predicted PGF of $N$ is  given by  \eqref{cards}, then the PGF of a random number of BMD models is 
\begin{align}
\Psi^{\mathrm{RU\mhyphen J\mhyphen BMD}}(h,g) = \mathrm G^{N^-}\!\big(\Psi^{\mathrm{BMD}}(h,g)\big).\end{align}
The nonzero probability events of this process have a random number of unlabeled targets and at most the same number of measurements. 

 \subsection{Unlabeled J-PDA and J-IPDA Targets} 
\label{JPDA/S}
Superposing a false alarm process with the U-J-BMD target measurement process gives the  U-J-PDA (unlabeled J-PDA) filter. It is shown in \cite{ASE2021} that  the     computational complexity of the  {U-J-PDA} intensity filter is $O(N\!M)$, where $N$ and $M$ are the numbers of targets and measurements, respectively.  This is significantly lower than the NP-hard complexity of the labeled  J-PDA filter.  The PGFL of the superposition of the  U-J-BMD target process whose PGFL is \eqref{SuperposedJPDAtheRealOne}  and an independent false alarm process with   PGFL \eqref{generalFA}  is   
 \begin{align}
\label{sjpda}
\boxed{
\Psi^{\mathrm{U\mhyphen J\mhyphen PDA}}(h,g)=\Psi^{\mathrm{FA}}(g) \big( \Psi^{\mathrm{BMD}}(h,g)\big)^N.
}
 \end{align}
Replacing $\Psi^{\mathrm{BMD}}(h,g)$ in  \eqref{sjpda} with the PGFL \eqref{BasicIBMD}  of the IBMD   yields the PGFL of the superposed U-J-IPDA (unlabeled J-IPDA) filter: 
 \begin{align}
\label{sjIpda}
\boxed{
\Psi^{\mathrm{U\mhyphen J\mhyphen IPDA}}(h,g)=\Psi^{\mathrm{FA}}(g) \big( \Psi^{\mathrm{IBMD}}(h,g)\big)^N.
}
 \end{align}
The $N^{\mathrm{th}}$ power of the  Bernoulli PGF  \eqref{BernoulliPGFAppdx} is the PGF of a binomial distribution, so the U-J-BMD and U-J-IPDA filters are  binomial intensity filters.   

With a Poisson false alarm process \eqref{PPP} with ${\lambda^{\mathrm{FA}}>0}$, the nonzero probability events of U-J-PDA have exactly $N$ unlabeled targets and any number of measurements.  The nonzero probability events for U-J-IPDA have up to $N$ targets and, with Poisson false alarms,  any number of measurements.

\subsection{Random Number of BMD Targets  (CPHD, PHD)} 
\label{CPHDandPHD}
We now assume that the number of targets $N$ in the U-J-PDA filter is random with a predicted PGF $ \mathrm G^{N^-}(z)$ given by \eqref{cards}.  Multiplying \eqref{sjpda} by the predicted probability  ${\Pr\{N^-\!=n\}}$ and summing $n$ from 0  to infinity  gives the PGFL of  the CPHD filter \cite{BakIvan}, \cite{2007Mahler},\cite{stochasticFlows}:
\begin{align}
\label{cphd}
\boxed{
\Psi^{\mathrm{CPHD}}(h,g)= \Psi^{\mathrm{FA}}(g) \, \mathrm G^{{N}^-}\!\big( \Psi^{\mathrm{BMD}}(h,g)\big).
}
\end{align}
The   PGF $ \mathrm G^{N^-}(z)$ and the  distribution $p^-(x)$ of the predicted target  process are inputs to the  filter. 

The U-J-PDA, IPDA, and  U-J-IPDA filters are special cases of the CPHD  filter.  To see this for the IPDA filter,   let  $\mathrm G^{N^-}\!(z)=1\!-\!\chi^{\!-}+\chi^{\!-}z$ in \eqref{cphd} and note that the result is identical to the PGF \eqref{IPDA}.  Similarly, for the  U-J-IPDA, let $N$ denote the  number of  targets, each of which may or may not exist. Assuming the targets are independent, the joint PGF is  $\mathrm G^{N^-}\!(z)=\left(1\!-\!\chi^{\!-}+\chi^{\!-}z\right)^N$. Substituting this expression into \eqref{sjIpda} gives the PGFL  of the J-IPDA filter  for  $N$ identical superposed targets (cf.   \eqref{gfJIPDA} for $N$ non-identical targets).  The U-J-PDA filetr is the special case $\chi^-=1$. 

Specifying Poisson PGFs   in \eqref{cphd} for the numbers of targets and false alarms gives the PGFL of the PHD filter \cite{BakIvan}, \cite{2003Mahler},\cite{stochasticFlows}:
\begin{align*}
    &\Psi^{\mathrm{PHD}}(h,g)\\
    & = \mathrm G^{\mathrm{Pois}(\lambda^{\mathrm{FA}})}\!\left(\textstyle\int_{\mathcal Y}g(y) p^{\mathrm{FA}}(y) \,\mathrm dy\right) \mathrm G^{\mathrm{Pois}(\lambda^{N^-}\!)}\big(\Psi^{\mathrm{BMD}}(h_{},g)\big).
\end{align*}
Explicitly, using  \eqref{PPP} and \eqref{BasicBMD}, the PGFL is  
\begin{equation}
\label{phd}
\boxed{
\begin{aligned}
    &\Psi^{\mathrm{PHD}}(h,g)= \exp\!\Big(\!\!-\!\lambda^{\mathrm{FA}}\!-\!\lambda^{N^-} \!\!+\lambda^{\mathrm{FA}} \textstyle \int_{\mathcal Y}g(y) p^{\mathrm{FA}}(y)\,\mathrm dy \bigg . \\
    & +\!\lambda^{N^-}\! \bigg . \!\!\int_{\!\mathcal X} \!h(x) p^{\!-}\!(x)\! \left(\!Qd(x)\!+\!P\!d(x)\!\textstyle\int_{\mathcal Y}  g(y) p(y|x)\,\mathrm dy\right)\!\mathrm dx\!\Big).
\end{aligned}
}
\end{equation}
The  nonzero probability events   may have any number of  targets  and any number of measurements.

The prior process of the PHD intensity filter is assumed to be a PPP.  Direct calculation (cf.  \eqref{BayesIsBeautiful})  shows that the  PGFL of the {\it{exact}} posterior  process is the product of the PGFLs of several independent  processes. One PGFL in the product  is the PPP of the expected number of targets conditioned on the target not being detected. The other PGFLs in the product are    Bernoulli point processes,  each  conditioned on one  measurement being the target measurement and the rest being false alarms.  Since PGFLs characterize point processes,  and the posterior PGFL is not the PGFL of a  PPP, it follows that the  Bayesian recursion is not closed.  

The PHD filter closes the recursion by approximating the exact posterior process with  a PPP whose  intensity is matched to the intensity  of the exact posterior. The posterior  intensity   can be found in either of two ways. The hard way is to compute the   derivative (cf.  \eqref{PGFL12}) of the posterior  PGFL.  The easy way is to recognize that the processes are independent, so the intensity function of the posterior  process is the sum of the intensity functions  of the constituent  processes.

\bigskip
{\color{black}
\section{Multisensor Unlabeled  Multitarget Tracking}
\label{UnlabeledMultisensorMultitarget}
}
The first subsection gives the multisensor version of the unlabeled  multitarget problems discussed in Section \ref{iFilter}.  The other subsection discusses a problem in which the measurement sets of multiple sensors are pooled, i.e., the sensor labels of the measurements are removed in an effort to anonymize the sensor origin. It  is   related to the difficult robotics problem called  SLAM (simultaneous localization and mapping) with DATMO (detection and tracking of moving objects) \cite{SlamDatmo}. 

\bigskip
\subsection{Random Number of Targets}

Most  versions of  multisensor unlabeled multitarget tracking problems assume that we know the number of sensors and the sensor labels of  the measurements, but that we do not know the measurement to target assignments. When there are $L$ conditionally independent sensors  and $N$ targets, the PGFL is $\Psi^{\mathrm{MS\mhyphen J\mhyphen BMD}}(h_{1:N},g_{1:L})$, which is given above by \eqref{GFBM1NmultiEll}. The sensors may be heterogeneous.  

Given the additional assumptions  detailed  in Section \ref{SuperBM} above, the targets can be superposed. Substituting $h_{n}=h$ into $\Psi^{\mathrm{MS\mhyphen J\mhyphen BMD}}(h_{1:N},g_{1:L})$ gives the PGFL for $N$ unlabeled targets and $L$ heterogeneous sensors as
\begin{equation}
\begin{aligned}
&\Psi^{\mathrm{MS\mhyphen UJ\mhyphen BMD}}(h,g_{1:L})  \\
&\qquad=\left( \int_{\mathcal X}  h(x) p^-(x) \prod_{\ell=1}^L \Psi^{\mathrm{BMD}|x}_{\ell}(g_\ell)\,\mathrm dx\right)^{\!\!N},
\end{aligned}
\end{equation}
where (cf. \eqref{bmdofnSensorEll}) 
$$\Psi^{\mathrm{BMD}|x}_{\ell}(g_\ell)=Qd^\ell(x)+P\!d^\ell(x)\!\!\int_{\mathcal Y_\ell} g_\ell(y) p_n^\ell(y|x)\,\mathrm dy.$$
The joint PGFL of the sensor false alarm processes, assuming they are mutually independent, is given by \eqref{JointFalseAla}.  The  multisensor unlabeled JPDA filter is derived from  the superposition of the false alarm processes and the unlabeled target processes, so the PGFL is 
\begin{equation}
\label{MSUJPDA}
\boxed{
\begin{aligned}
    &\Psi^{\mathrm{MS\mhyphen UJ\mhyphen PDA}}(h,g_{1:L}) \\
    &\quad =\Psi^{\mathrm{FA}}_{1:L}(g_{1:L})\left( \int_{\mathcal X}  h(x) p^-(x) \prod_{\ell=1}^L \Psi^{\mathrm{BMD}|x}_{\ell}(g_\ell)\,\mathrm dx\right)^{\!\!N}.
\end{aligned}
}
\end{equation}
The PGFL of the multisensor CPHD filter is the natural extension of this PGFL that assumes that the number of targets is random with PGFL given by \eqref{cards}.  The PGFL is 
\begin{equation}
\label{MSUCPHD}
\boxed{
\begin{aligned}
    &\Psi^{\mathrm{MS\mhyphen UJ\mhyphen CPHD}}(h,g_{1:L}) \\
    &\,\,=\Psi^{\mathrm{FA}}_{1:L}(g_{1:L})\,\mathrm G^{N^-}\!\!\left( \int_{\mathcal X} \! h(x) p^-(x) \prod_{\ell=1}^L \Psi^{\mathrm{BMD}|x}_{\ell}(g_\ell)\,\mathrm dx\right)
\end{aligned}
}
\end{equation}
corresponds to a random number of unlabeled targets with $L$ heterogeneous sensors.  The multisensor PHD intensity filter is derived from the special case with Poisson numbers of false alarms and targets.   

{\color{black}
\subsection{Random Numbers of Targets and Sensors} 
\label{DoublePoisson}
}
The PGFL  \eqref{MSUCPHD} includes two terms that are  products over $L$ sensors. One product is for  sensor false alarms, and the other is for the conditional target measurement processes.   This section considers the intriguing possibility of replacing a fixed number of known sensors with a random number of unlabeled sensors in a manner that is analogous in the switch from labeled target to unlabeled targets. The practicality of this problem is an open question. 

The multisensor J-BMD problem discussed in Section \ref{MultiSensorBMD}   assumes  that the sensor states (e.g., locations, orientations, etc.) are known, and that the  sensor measurement sets  are labeled, i.e., it is known which sensor generated which measurement. Both assumptions are dropped  in this section.

The problem considered here assumes that each sensor collects its own set of measurements, including false alarms, and reports them to a central site. The central site anonymizes the measurements by removing the sensor labels and pooling the measurements.    Pooling the measurements also removes knowledge of   the number of sensors. The sensors  are assumed to be homogeneous and also independent, conditioned on target states, and to have a common frame of reference, i.e., there  are no sensor-to-sensor measurement biases. The  problem is to simultaneously estimate   the numbers and distributions of  targets and sensors given  measurements that lack  target and sensor labels.  


The state of a sensor  is a  point in the (Euclidean) state space $\mathcal S$.  We model the unknown number and states of  unlabeled targets as a finite point process on $\mathcal X$, so we also model the number and states of the unlabeled sensors 
as a finite point process on $\mathcal S$.  The point process model implicitly   assumes  that the sensor motion models are identical, which is true also of  the  target motion models.

Denote the number of sensors at the previous scan by $L$.  We assume that $L$  is a random nonnegative integer whose PGF  from  the previous scan, $\mathrm G^L(z)$,  is predicted forward to the current scan:   
\begin{align}
\label{Sensorcards}
     \mathrm G^{L^-}(z) = \textstyle \sum_{\ell=0}^\infty \Pr\{L^-\!=\ell\}\,z^\ell,
\end{align}
where $z$ is the indeterminate variable. The sensor state is modeled as a  Markovian process on  $\mathcal S$. Its motion from the previous state $s_{k-1}\in\mathcal S$ to the current state $s_k\in\mathcal S$ is  (cf. \eqref{SIR}) 
\begin{align}
    \label{SIRR}
p_k^{\!-}(s_k|\mathbb Y_{k-1})=\textstyle\int_{\mathcal S}p_{k}(s_k|s_{k-1}) p_{k-1}(s_{k-1}|\mathbb Y_{k-1}) \,\mathrm ds_{k-1},
\end{align}
where $p_{k-1}(s_{k-1}|\mathbb Y_{k-1})\equiv p_k^{\!-}(s_k)$ is the  prior and $p_{k}(s_k|s_{k-1})$ is the  motion model. As with targets, we drop the time subscripts and use (reluctantly) the function argument  to distinguish the sensor functions from the analogous target functions. The sensor indeterminate function is ${f:\mathcal S\rightarrow \mathbb C}$. The pooled measurement set is assumed to be superposed with an independent false alarm process, so the PGF of the ``forensic'' problem  is 
\begin{align*}
    &\Psi^{\mathrm{Forensic}}(h_{},g,f)=\Psi^{\mathrm{FA}}(g) \\
    &\,\times\!\mathrm G^{{N}^-}\!\Big(\!\textstyle\int_{\mathcal X} h(x) p^-(x)\,\mathrm G^{{L^-}}\!\Big(\!\textstyle \int_{\mathcal S}  f(s) p^-(s)\,\Psi^{\mathrm{BMD}|x,s}(g)\,\mathrm ds\Big)\mathrm dx\Big),
\end{align*}
where  the conditional measurement PGFL is defined by (cf. \eqref{BMDgf})  
\begin{align*}
    \Psi^{\mathrm{BMD}|x,s}(g)\equiv Qd(x,s)+P\!d(x,s)\textstyle\int_{\mathcal Y}  g(y)\, p(y|x,s)\,\mathrm dy, 
\end{align*}
where  $P\!d(x,s)$ is the probability of detecting a target  at $x\in\mathcal X$ when the sensor is at $s\in\mathcal S$, and ${Qd(x,s)=1\!-\!P\!d(x,s)}$.   Defining the expression  
\begin{align*}
    \Psi^{\mathrm{BMD}|x}(f,g)=\textstyle\int_{\mathcal S} f(s) p^-(s)\,\Psi^{\mathrm{BMD}|x,s}(g)\,\mathrm ds, 
\end{align*}
the forensic PGFL   takes the   more compact form 
\begin{equation}
\label{slammer}
\boxed{
\begin{aligned}
    &\Psi^{\mathrm{Forensic}}(h_{},g,f)  \\
    &\,\,\, = \Psi^{\mathrm{FA}}(g)\, \mathrm G^{{N}^-}\!\Big(\!\textstyle\int_{\mathcal X} h(x) p^-(x)\,\mathrm G^{{L^-}}\!\big(\textstyle \Psi^{\mathrm{BMD}|x}(f,g)\,\mathrm dx\Big).
\end{aligned}
}
\end{equation}
By comparing \eqref{slammer} to   \eqref{cphd},  it is seen that  the relationship between sensors and measurements in the PGFL of  the forensic problem  is,  conditioned on target state $x$,  analogous to the relationship   between targets and measurements in the PGFL for  CPHD.    

If  the predicted point  processes for targets and sensors are Poisson, i.e., if $\mathrm G^{N^-}(\cdot)$ and $\mathrm G^{L^-}(\cdot)$ are Poisson distributed, then  the Bayesian  forensic filter  recursion can be closed by approximating the exact posterior process using the intensity functions of two marginal processes, one for sensors and the other for targets, both conditioned on the same measurement set. The PGFL of the marginal sensor-measurement process is  $\Psi^{\mathrm{Forensic}}(1,g,f)$, and the PGFL of the marginal  target-measurement process is $\Psi^{\mathrm{Forensic}}(h,g,1)$.  



\bigskip
{\color{black}
\section{Hybrid  Labeled and Unlabeled Targets}
\label{LabeledAndUnlabeled}
}
The problems in this section combine aspects of both PDA and PHD intensity filters. Several single target problems are discussed. Multisensor  hybrid problems  are outside the scope of the present paper.  

The problem in the first subsection is closely related to ``one scan deep''  MHT filters. It assumes there are $N$ labeled targets and an unknown number of unlabeled targets.  

The next  subsection   concerns tracking one ``target of interest'' in a swarm of targets.  The swarm is a mixture of known targets and other targets that may or may not be reliably detected. The target of interest is labeled, and   the targets in the swarm are unlabeled.  The problem arises when the target swarm is so large that it is impractical to track all the targets  just to hold track on the one target of interest.   

A related problem is to  track a specified  number of  targets, each of which is observed by the sensor as a ``cloud'' of  measurements of  what we call target highlights. Alternatively,  a target cloud can  be interpreted  as several  individual targets moving  in a coordinated manner, e.g., in group formation. The  filter  estimates a  highlight intensity function for each target   from which, in some applications, target shape and orientation could be estimated.  

It is shown in the  final subsection  that the PGFL for  one or more point targets can be converted into the PGFL for the same problem but with extended targets.

\bigskip
\subsection{J-PDA with Unlabeled Targets}
\label{JIPDAinaCloud}
The J-PDA filter assumes that there are exactly $N$ targets, each labeled. In addition to these $N$ targets we now assume there is an unknown number of independent, unlabeled targets. The unlabeled target process can be interpreted as a model for targets that are not yet  detected. The $N$ labeled targets may have different state spaces, but unlabeled targets are assumed to have the same state space $\mathcal X$.  The product of the PGFLs for false alarms, for the $N$ labeled targets of J-BMD  \eqref{GFBM1Nmulti},   and for the unlabeled targets of  CPHD (cf. \eqref{cphd}) gives the joint PGFL 
\begin{equation}
\label{JPDAinaCloud}
\boxed{\begin{aligned}
    &\Psi^{\mathrm{J\mhyphen U\mhyphen PDA^{}}}(h_{1:N},h,g)\\
    &\quad =\Psi^{\mathrm{FA}}(g) \left(\prod_{n=1}^N \Psi^{\mathrm{BMD}}_n(h_n,g) \right) \mathrm G^{{N}^-}\!\big( \Psi^{\mathrm{BMD}}(h,g)\big).
\end{aligned}
}
\end{equation}
The BMD filter in the argument of $ \mathrm G^{{N}^-}(\cdot)$ is identical to \eqref{BasicBMD}.   

Given the false alarm process,  the inputs to the U-J-PDA$^{}$ filter are parameterized by the $N$ predicted pdfs   $p_n^-(x)$ of the labeled targets, together with the parameters of the  unlabeled target process. If the unlabeled target process is modeled as a PPP, then it is parameterized by the  expected number of unlabeled targets,  $\lambda^{\mathrm{ukn}}$, and the predicted point process pdf, $p^{\mathrm{ukn}^-}(x)$. The factors in the PGFL \eqref{JPDAinaCloud} show that  it is close kin to the  single-scan MHT filter. 

\subsection{PDA in a Swarm of Unlabeled   Targets}
\label{toiJIPDAinaCloud}
The problem is to track a single  target of interest (TOI) in a heterogeneous ``swarm'' of other targets and false alarms.  It is assumed that the swarm comprises two kinds of targets, namely, a given number of  known  targets and  a random number of  unknown  targets. We do not track  targets in the swarm.  Consequently,  the known target process is  an U-J-PDA process with $N$ superposed (i.e., unlabeled) targets (cf. \eqref{sjpda}) on the space  $\mathcal X_\mathrm{kwn}$.  We  model the unknown target process as a finite point process on $\mathcal X_\mathrm{ukn}$.   The target state spaces need not be the same, and both may differ from the state space of the TOI, which is denoted $\mathcal X_\mathrm{toi}$. All targets, including the TOI, are modeled as BMD  processes.   All target  measurements are   in the space   $\mathcal Y$.  Target measurements  are superposed with an independent false alarm finite point process on  $\mathcal Y$.

Given this information, we cobble together the PGFL by multiplying the PGFLs of several filters already presented:
\begin{equation}\label{PDAswarm}
\boxed{
\begin{aligned}
    \Psi^{\mathrm{Swarm}}&(h_{\mathrm{toi}},h_{\mathrm{kwn}},h_{\mathrm{ukn}},g)=\Psi^{\mathrm{FA}}(g)\,\Psi^{\mathrm{BMD}}\big(h_{\mathrm{toi}},g\big)\\
    &\times  \big(\Psi^{\mathrm{BMD}}(h_{\mathrm{kwn}},g)\big)^{\!N} \, \mathrm G^{\mathrm{Pois}(\lambda)}\big(\Psi^{\mathrm{BMD}}(h_{\mathrm{ukn}},g)\big).
\end{aligned}
}
\end{equation}
The  false alarm process  $\Psi^{\mathrm{FA}}(g)$ is often assumed to be Poisson, but  it  can be arbitrarily specified. The  terms  $\Psi^{\mathrm{BMD}}(h_{\mathrm{toi}},g)$  and $\big(\Psi^{\mathrm{BMD}}(h_{\mathrm{kwn}},g)\big)^{\!N}$  model the measurement process of the TOI and the (superposed) measurement process of   the $N$ known targets, respectively. The last term,  $\mathrm G^{\mathrm{Pois}(\lambda)}\big(\Psi^{\mathrm{BMD}}(h_{\mathrm{ukn}},g)\big)$, where $ \mathrm G^{\mathrm{Pois}(\lambda)}(\cdot)$ is the PGF \eqref{PoissonGF} of the  Poisson distribution,     models the measurement process for an unknown number of targets. It need not be Poisson and  can, if desired, be arbitrarily specified.       

The inputs to the PDA swarm filter comprise:
\begin{enumerate}
    \item the predicted prior pdf of the TOI,  $p_{\mathrm{toi}}^-(x)$, $x\in\mathcal X_\mathrm{toi}$,
    \item the predicted  pdf of the  U-J-PDA point process with $N$ targets, $p_{\mathrm{kwn}}^-(x)$, $x\in\mathcal X_\mathrm{kwn}$, 
    \item the predicted intensity function of the unknown target process. 
\end{enumerate}  
The third input  is more involved if the Poisson model for the  unknown, i.e., not yet detected,  targets  is changed to the general model $\mathrm G^{{N}^-}\big( \Psi^{\mathrm{BMD}}(h_{\mathrm{ukn}},g)\big)$.

\subsection{J-PDA with  Target Highlights  (JiFi)}
\label{JiFi}   JiFi (\underline J-PDA \underline intensity \underline{Fi}lter) is the J-PDA filter for targets that the sensor observes  as a collection of one or more points  \cite{JiFi}.  We call these points  highlights.  In high resolution radar applications, for example, the highlights can be the specular reflections of a target that are observed by the sensor. Such reflections will fluctuate in number and location from scan to scan due to changing  target-sensor orientations. 

JiFi, like the J-PDA filter,  assumes that exactly $N$ targets are known. The   highlight process of target $n$ is assumed to be a finite point processes on the state space, $\mathcal X_n$. The PGF of the random number of highlights for target $n$ can be arbitrary, but we assume it is Poisson (cf. \eqref{PoissonGF}) and  the predicted expected number of highlights is   $\lambda_n^-$.  With these assumptions, the PGFL of the highlight process for target $n$ is 
\begin{align}
    \label{highlites}
    \Psi^{\mathrm{HiLi}}(h_n,g)=\mathrm G^{\mathrm{Pois}(\lambda_n^-)}\big(\Psi_n^{\mathrm{BMD}}(h_{n},g)\big),
    \end{align}  
where $\Psi_n^{\mathrm{BMD}}(h_{n},g)$ is given by \eqref{BasicBMDn}.  Target highlight processes are assumed independent of each other and of the false alarm process,  so the overall  PGFL  for Poisson JiFi  is 
\begin{equation}
\boxed{
    \label{JiFiPGF}
    \Psi^{\mathrm{JiFi}}(h_{1:N},g)=\Psi^{\mathrm{FA}}(g)\prod_{n=1}^N \mathrm G^{\mathrm{Pois}(\lambda_n^-)}\big(\Psi_n^{\mathrm{BMD}}(h_{n},g)\big).
}
\end{equation}
The posterior process  on ${\mathcal X_1\!\times\!\cdots\!\times\!\mathcal X_N}$ is not  Poisson, but its joint intensity function factors into a product of intensity functions on $\mathcal X_n$, one for each target highlight process. The factorization is an exact analytical result, not a mean field  approximation.  To close the recursion, we approximate the posterior process  on ${\mathcal X_1\!\times\!\cdots\!\times\!\mathcal X_N}$ as a product of PPPs whose intensities are those of the posterior process. Poisson JiFi has low computational  complexity \cite{JiFi}, which  is remarkable considering the high  complexity of J-PDA.  



\subsection{General Extended Targets}
\label{extendedT}
Targets that obey the ``at most one measurement per target'' rule  are called point targets, and those that do not are called extended targets. The PGFLs of many  filters for point targets generalize  easily  to  extended targets that appear in the sensor as a random number of measurements.  Denote the PGF of the number $M$ of  measurements generated by the sensor due to a target at $x$ by    
\begin{align}
    \label{PGFmeasuremetnNo}
    \mathrm G^{M|x}(z)=\textstyle\sum_{m=0}^\infty \Pr\{M=m|x_{}\}\, z^m.
\end{align} 
The PGFL for the BMD problem for an extended target  is     
\begin{align}\label{BMDgffffff}
    \Psi^{e\mathrm{BMD}}(h,g)=\textstyle\int_{\mathcal X} h(x) p^-(x) \Psi^{e\mathrm{BMD}|x}(g)\,\mathrm dx,
\end{align}
where   
\begin{equation}
\label{BMDgffff}
    \Psi^{e\mathrm{BMD}|x}(g)=\mathrm G^{M|x}\Big(\textstyle\int_{\mathcal Y} g(y) p(y|x)\,\mathrm dy\Big).
\end{equation}
The  PGFL of the BMD filter for point targets \eqref{BasicBMD} is the special case  $\mathrm G^{M|x}(z)=Qd(x)+P\!d(x)\, z$.  

Point target problems  formulated in terms of PGFLs of BMD targets can be generalized to extended targets simply by replacing  BMD target PGFLs with  PGFLs of the form $\Psi^{\mathrm{eBMD}|x}(g)$.  
This  method is applicable to both labeled and unlabeled target problems.  A prominent example of this is the JiFi tracker of Section \ref{JiFi}.  It starts with the JPDA filter for point targets \eqref{gfJPDA} and replaces $\Psi^{\mathrm{ BMD}}(h_{n},g)$ with the Poisson  highlight  model  \eqref{highlites}
$$
\Psi^{e\mathrm{BMD}}(h_n,g)\equiv \Psi^{\mathrm{HiLi}}(h_n,g).
$$
Other examples are left to the reader.

\bigskip
{\color{black}
\section{Trajectory Filters for  Labeled and Unlabeled Multitarget Tracking}
\label{MultipleScan}
}

A  target trajectory begins and ends at an unknown time and state.   An unknown number of targets may be present, each  with unknown endpoints in time and state.  From beginning to end, a sensor may generate detections (i.e., measurements) of some of the targets some of the time, but the measurement assignments are unknown. Superposed with the target measurements, if any, is a false alarm process.  A sensor  gathers a time sequence of scan measurements over some period of time  and sends it all to the tracking filter, whose job is to make sense of it all. 

Trajectory filters are well studied for   labeled targets, where they are better known as interval or smoothing filters. They are  largely unexplored for unlabeled targets. We seek a filter that estimates the number of targets  present at each time in the sequence, and a state estimate  for  the targets that are present at that time.  

This section begins by deriving the PGFL for the set of trajectories that is generated by one target that is initialized at time $k=0$. The PGFL is a found by a backward recursion.   The trajectory process is a branching process.  Most of the  mathematical literature on branching processes is devoted to  asymptotic results, and to what is termed ``extinction'' probabilities for  ``perfectly observed'' processes. 

Our trajectory model is distinguished in that it is a state-dependent and time-inhomogeneous branching process over a finite time interval, and it is observed in a noisy environment.  Our goal is a Bayesian analysis of the trajectories over the observation interval, not {\it{a priori}} extinction probabilities. Many of the Bayesian aspects of our model appear to be new, but  many variations of branching processes have been studied for  many different applications for over 100 years, so our claim to novelty is tentative.  

The one target trajectory PGFL  is the basis for deriving the PGFLs for  interval filters for both labeled and unlabeled targets  \cite{RLS2017}. The conditional PGF of the number of  trajectories  appears to be new.  

Multisensor interval problems are not considered here. 
The relevant PGFLs are relatively straightforward to derive by  modifying the derivation in this section.

\bigskip
\subsection{PGFL for Trajectories Initialized by One Target}
\label{deriatioSmoothing}
The scans in the sequence, or batch, of measurements, are indexed by $k\in\{1,2,\ldots,K\}$. The index of the current scan is $K$. The index of the scan prior to the first scan is ${k=0}$. The random target process at time $k$ is denoted by $X_k$, and its state space is $\mathcal X_k$.  
The measurement processes have realizations in the same space, $\mathcal Y$.  

Let $x_k$ denote a point in $\mathcal X_k$ at time ${k,\,\,0\leqslant k\leqslant K}$. We assume   that a target at $x_{k-1}$ transitions to $x_k$ at time $k$ according to the transition pdf $p_k(x_k|x_{k-1})$. A  target detection process for one target at $x_k\in\mathcal X_k$ generates a random number of  measurements.  The conditional PGF at scan $k$ is 
\begin{align}
    \label{measuremetnNo}
    \mathrm G^{M_k|x_k}(z)=\textstyle\sum_{m=0}^\infty \Pr\{M_k=m|x_{k}\}\, z^m,\quad 1\leqslant k\leqslant K.
\end{align} 
Targets that  generate at most one measurement per scan is the special case   
\begin{align}
\label{simplezz}
    \mathrm G^{M_k|x_k}(z)=Qd_k(x_k)+P\!d_k(x_k)\,z,
    \end{align}
where the probability of detecting a target at $x_k$ at time $k$ is $P\!d_k(x_k)$. 

A target at $x_{k-1}$ at time ${k-1}$  may ``branch''   into a random number of targets, $B_k$,  at time $k$. The   conditional PGF of $B_k$ is   
\begin{align}
    \label{targetNo}
    \mathrm G^{B_k|x_{k-1}}(z)=\textstyle\sum_{b=0}^\infty \Pr\{B_k=b
    \,|
    \,x_{k-1}\}\, z^{b},\,\, 1\leqslant  k\leqslant K.
\end{align} 
The coefficient   $\Pr\{B_k=0\,|
\,x_{k-1}\}$ is the probability that a  target terminates in state $x_{k-1}$ at time $k\!-\!1$.  The  special  case    
\begin{align}   
\label{simpleGz}
\mathrm G^{B_k|x_{k-1}}(z)=Qs_k(x_{k-1})+Ps_k(x_{k-1}) \,z 
\end{align}
is a birth-death process in which the target successfully    transitions to time $k$   with probability 
$$
{Ps_k(x_{k-1})=\Pr\{B_k=1|x_{k-1}\}}
$$
and is unsuccessful    with probability    ${Qs_k(x_{k-1})=1-Ps_k(x_{k-1})}$. More generally, if   \eqref{targetNo}  has quadratic or higher order terms, the number of targets can cascade and grow over time.


The indeterminate functions  $h_{1:K}$ and $g_{1:K}$  correspond, respectively,  to the  random processes $X_{1:K}$ and $Y_{1:K}$.   The  PGFL of the  joint sequence $(X_{1:K},Y_{1:K})$ is denoted by $\Psi^{N_{1:K}\!M_{1:K}}(h_{1:K},g_{1:K})$.  It is computed by backward recursion. 

The PGFL of one measurement in $y\in \mathcal Y$ conditioned on one target at $x_K\in\mathcal X_K$ at time $K$ is 
$$
\textstyle{\int_{\mathcal Y} g_K(y) p_K(y|x_K)\, \mathrm dy,}
$$
where $p_K(y|x_K)$ is the measurement likelihood function.  The PGFL for   ${m_K\geqslant 0}$ i.i.d.  measurements conditioned on one  target at $x_K$ is $$\textstyle{\big(\int_{\mathcal Y} g_K(y) p_K(y|x_K)\, \mathrm dy\big)^{m_K}.}$$  Multiplying by ${\Pr\{M_K=m_K|x_{K}\}}$,  summing over all ${m_K\geqslant 0}$, and  the PGF \eqref{measuremetnNo} gives the PGFL of the number of measurements in scan $K$ conditioned on one target at $x_K$ as
\begin{align}
    \label{rec1}
 \Psi^{M_{\!K}|x_K}(g_K)\equiv \mathrm G^{M_{\!K}|x_K}\Big(\textstyle\int_{\mathcal Y} g_K(y) p_K(y|x_K)\, \mathrm dy\Big).
 \end{align}
Each of the ${B_K=b_K\geqslant 0}$ targets generates a measurement process  with the PGFL  \eqref{rec1}.  The $b_K$ targets are independent, by assumption, so  their measurement processes are independent. The  PGF of the measurements generated by   $b_K$ i.i.d. targets   is 
\begin{equation} 
\label{recursivestep}
\begin{split}  
\Big(\textstyle \int_{\mathcal X_K} h_K(x_K) p_K(x_K|x_{K-1})\, \Psi^{M_{\!K}|x_K}(g_K)\,\mathrm dx_K\Big)^{b_K}.  
\end{split}
\end{equation}
This PGF is conditioned on the presence of  one target at ${x_{K-1}\in\mathcal X}$ at time  ${K\!-\!1}$.    Multiplying by ${\Pr\{B_K=b_K|x_{K-1}\}}$,  summing over  $b_K\geqslant 0$, and  the PGF \eqref{targetNo} gives the PGFL of a random number of target-measurement processes in scan $K$ conditioned one  target at $x_{K-1}$ in the previous scan: 
\begin{equation}
    \label{sidestep}
\begin{split}
    \begin{aligned}
    &\!\Psi^{N_{\!K}\!M_{\!K}|x_{K\!-\!1}}(h_K,g_K) \\
    & \!\!= \mathrm G^{B_{\!K}|x_{K\!-\!1}}\!\Big(\textstyle \!\int_{\!\mathcal X_K} \!\!h_K(x_K) p_K(x_K|x_{K\!-\!1})\Psi^{M_{\!K}|x_K}(g_K)\,\mathrm dx_K\!\Big).
\end{aligned}
\end{split}
\end{equation}
The first step of the backward recursion is now in sight. 

The  joint PGFL of the  target-measurement process at time $K$ {\it{and}}  the measurement process at time ${K\!-\!1}$ is  the product
\begin{align}
\label{step}
    \Psi^{N_{\!K}\!M_{\!K}|x_{K\!-\!1}}(h_K,g_K) \,\Psi^{M_{K\!-\!1}|x_{K\!-\!1}}(g_{K\!-\!1}), 
\end{align}
where $\Psi^{M_{K\!-\!1}|x_{K\!-\!1}}(g_{K\!-\!1})$ is PGFL given by \eqref{rec1} with $K$ replaced by $K-1$. It is conditioned on the presence of one target  at time $K\!-\!1$ in state $x_{K-1}$. The PGFL of  $B_{k-1}=b_{K-1}\geqslant 0$ i.i.d.  such processes is 
$$\Big(\Psi^{N_{\!K}\!M_{\!K}|x_{K\!-\!1}}(h_K,g_K) \,\Psi^{M_{K\!-\!1}|x_{K\!-\!1}}(g_{K\!-\!1})\Big)^{b_{K-1}}.$$
Multiplying   by the branching probability  ${\Pr\{B_{K-1}\!= b_{K-1}|x_{K-2}\}},$ summing over  $b_{K-1}$, and  using the PGF \eqref{targetNo} gives the  PGFL of the  target-measurement process  at time  $K$ {\it{and}} the  target-measurement process  at time $K\!-\!1$, 
\begin{align}
    \label{recurs2}
    \begin{split} &\Psi^{N_{\!K\!-\!1}\!M_{\!K\!-\!1}|x_{K\!-\!2}}(h_{K\!-\!1:K},g_{K\!-\!1:K})\\
     &\quad= \mathrm G^{B_{K\!-\!1}|x_{K\!-\!2}}\Big(\textstyle \!\int_{\mathcal X} h_{K\!-\!1}(x_{K\!-\!1})\, p_{K-1}(x_{K\!-\!1}|x_{K\!-\!2})\\
     &\quad\,\,\,\quad\times \Psi^{N_{\!K}\!M_{\!K}|x_{K\!-\!1}}(h_K,g_K)\,\Psi^{M_{K\!-\!1}|x_{K\!-\!1}}(g_{K\!-\!1})\,\mathrm dx_{K\!-\!1}\!\Big).
     \end{split}
\end{align}
This PGFL is a   function of ${h_{K-1}}$ and ${g_{K-1}}$ at scan ${K\!-\!1}$ and ${h_{K}}$ and ${g_{K}}$  at scan $K$.  Proceeding in this way gives the following  recursion.

\begin{equation*}
    \label{recursion}
\boxed{
\begin{aligned}
        &{\text{RECURSION: PGFL of  Trajectories Initialized by One Target}}\\
        &\bullet{\text{Initialize: for all $x_K\!\in\!\mathcal X_K$, define}}\\
        &\qquad \psi^{{N}_{\!K\!+\!1}\!M_{\!K\!+\!1}|x_{K}}\big(h_{K+1},g_{K+1}\big)\equiv 1\\
        &\bullet{\text{FOR: $k=K$ to $k=1$ in steps of $-1$}} \\
        &\,\,\,\,\,\, \text{Compute, for all $x_{k-1}\in \mathcal X_{k-1}$,}\\
        &\qquad \psi^{N_{k}M_{k}|x_{k\!-\!1}}\big(h_{k},g_{k}\big)\\
        &\qquad\quad= \mathrm G^{B_{k}|x_{k\!-\!1}}\!{\color{black}\bigg(} \int_{\mathcal X_k} h_k(x_k)\, p_k(x_k|x_{k\!-\!1})\\
        &\quad\quad\quad\quad\quad\quad\times \psi^{N_{\!k+1}\!M_{\!k+1}|x_{k}}\big(h_{k+1},g_{k+1}\big)\\
        &\quad\quad\quad\quad\quad\quad\qquad\times \mathrm  G^{M_{\!k}|x_k}\Big(\textstyle{\int_{\mathcal Y}} \,g_k(y) p_k(y|x_k)\, \mathrm dy\Big)\,\mathrm dx_k{\color{black}\bigg)}\\
        &\bullet{\text{END FOR}}\\
        &\bullet{\text{Compute the PGFL for all single-target trajectories:}}\\
       &\qquad\Psi^{N_{1:K}\!M_{1:K}}(h_{1:K},g_{1:K})\\
       &\,\,\qquad=\!\int_{\!\mathcal X_0} p_0(x_0)\,  \psi^{N_{1}\!M_{1}|x_{0}}(h_{1},g_{1}) \,\mathrm dx_0\\
       &\text{END RECURSION}\\
    \end{aligned}
}
\end{equation*}

\noindent In the last step, $p_0(x_0)$ is the prior pdf  of the initializing target  at time ${k=0}$.  (In sliding batch implementations, $p_0(x_0)$ is conditioned on  data in earlier scans.) It is easily verified that the recursion  gives $\Psi^{N_{1:K}M_{1:K}}(h_{1:K},g_{1:K})\equiv 1$ for $h_{1:K}(\cdot)=1$ and $g_{1:K}(\cdot)=1$.  

Realizations of  one  trajectory-measurement process are highly variable. For example, if the detection and branching processes  are given by \eqref{simplezz} and \eqref{simpleGz}, respectively, a realization may comprise  a sequence of singleton   measurement sets that begins at time $k'$ and ends at time $k''$, where ${1 \leqslant k'\leqslant k'' \leqslant K}$.  If the detection probability is less than one, a realization might also  comprise several such segments.

It is assumed that there is a  false alarm process at each time in the interval. These processes are assumed independent of each other, so the joint PGFL of the false alarm  process is 
\begin{align}
    \label{FAforInterval}
    \Psi^{\mathrm{FA}}(g_{1:K})=\textstyle\prod_{k=1}^K \Psi_k^{\mathrm{FA}}(g_{k}),
\end{align} 
where $\Psi_k^{\mathrm{FA}}(g_{k})$ is the false alarm process at time $k$.  

If a new and  unrelated target enters the scene at time ${k_0\geqslant 1}$, it would   generate a set of  trajectories from time $k_0$  to time $K$. The PGFL  of  such an ``immigrant'' target process would be computed by the same recursion, but  starting at time $k_0$. The recursion would  use immigrant-specific  motion models and, assuming independence, it would use the indeterminate functions $h_{k_0,K}$.  
The  PGFL of the superposition of the  trajectory processes  is the product of their PGFLs. Immigrant targets are  considered in \cite{RLS2017} but are not discussed further here.

\subsection{PGFL for a Known Number of  Targets}
The  PGFL for the one-target interval PDA filter is a special case of the general recursion.  To see this, note that the PGF ${\mathrm G^{B_k|x_{k-1}}(z)=z}$  models the assumption that one target is always present at each time $k$. Computing the recursion with these PGFs and multiplying the output by the false alarm process \eqref{FAforInterval} yields the PGFL for the one-target interval PDA filter. The ``at most one target is present'' interval  IPDA filter is  similar  but uses the branching PGF  ${\mathrm G^{B_k|x_{k-1}}(z)=1-\chi(x_{k-1}) + \chi(x_{k-1}) z}$.  
   

If  trajectories are initialized by $N$ known mutually independent and labeled targets, then there are $N$ labeled   trajectories, each with a PGFL of the form given by the recursion. These PGFLs can differ in many ways, but assuming for simplicity that they have the same target state spaces, motion models, etc., their PGFLs    differ in only one way --- they  have  target-specific   indeterminate functions, $h^n_{1:K}$.   The target-specific PGFLs do, however, use the same measurement indeterminate functions, $g_{1:K}$. The trajectory processes are independent of each other and of the false alarm process, so the joint PGFL is the product
$$\Psi^{N_{1:K}\!M_{1:K}}\big(h^{1:N}_{1:K},g_{1:K}\big)\!=\!\Psi^{\mathrm{FA}}(g_{1:K})\!\prod_{n=1}^N\!\Psi^{N_{1:K}\!M_{1:K}}\big(h^n_{1:K},g_{1:K}\big).
$$
This PGFL for labeled targets is very closely related to $K$-scan MHT filter.   A careful study of the  relationship remains to be done.

As we did to find the PGFL of the  U-J-PDA filter in Sect. \ref{JPDA/S}, we can do here too:  Remove the target labels and superpose the $N$ target trajectories   by setting $h^n_{1:K}=h_{1:K}$ for all  $n$.  This gives the PGFL of the  interval U-J-PDA filter:   
$$
\Psi^{N_{1:K}\!M_{1:K}}\big(h^{}_{1:K},g_{1:K}\big)\!=\!\Psi^{\mathrm{FA}}(g_{1:K})\Big(\Psi^{N_{1:K}\!M_{1:K}}\big(h_{1:K},g_{1:K}\big)\Big)^{\!N}\!\!.
$$
The computational complexity  of this filter is unknown. Since  superposing the targets in the J-PDA filter leads to a low complexity filter, namely the U-J-PDA filter,  it is not unreasonable to think that the superposition of interval J-PDA filters will also be a low complexity filter.   The possibility remains to be studied.

\subsection{PGFL for a Random Number of  Unlabeled Targets}
\label{RecursionStatement}
The PGFL for a random number of trajectories is derived from the PGFL for one trajectory in a manner that is similar to how, in earlier sections, multitarget PGFLs are derived from single target PGFLs. 
Denote  the number of targets at time $k=0$  by $N_0$ and its PGF by  
\begin{align}
\label{cardsSmoothing}
     \mathrm G^{N_0}(z) = \textstyle \sum_{n=0}^\infty \Pr\{N_0\!=n\}\,z^n.
\end{align}
The PGFL of the trajectories generated by one target at time $k=0$ is given by the recursion. The  PGFL of the    trajectories generated by $N_0$ targets at time $k=0$ is 
\begin{align}
    \label{smoothe}
\Psi^{\mathrm {Traj}}(h_{1:K},g_{1:K})\equiv \mathrm G^{N_0}\Big(\Psi^{N_{1:K}M_{1:K}}\big(h_{1:K},g_{1:K}\big)\Big).
\end{align}
The PGFL of the false alarm process at time  $k$ is $\Psi_k^{\mathrm{FA}}(g_k)$. Assuming the false alarms processes  are independent of each other and of the  trajectory  process, the PGFL of the superposition of the target trajectory and false alarm processes is 
\begin{align}
    \label{smootherer}
\boxed{    \Psi^{\mathrm{Batch}}(h_{1:K},g_{1:K})=\Psi^{\mathrm{FA}}(g_{1:K})\,\Psi^{\mathrm {Traj}}(h_{1:K},g_{1:K}). 
}
\end{align}
The marginal PGFL for the interval filter at time $k$ is found by setting $h_{k'}(x)=1,\, k'\ne k$.  In particular, marginalizing the joint PGFL over  trajectory histories up to but not including time $K$ gives 
\begin{align}\label{endALL}
  \begin{split}  \Psi^{\mathrm{Batch}}(h_{K},g_{1:K})
    \equiv\left.\Psi^{\mathrm{Batch}}(h_{1:K},g_{1:K})\right|_{\substack{h_k(x_k)=1\\k=1,\ldots,K-1}},
    \end{split}
\end{align}
Given a sequence of measurements $\mathbb Y_1,\ldots,\mathbb Y_K$, the PGFL of the posterior process is the (appropriately defined)  normalized mixed derivative of this PGFL.  The intensity  of the posterior  process at time $K$ is the expected number of trajectories whose endpoints are at     $x\in\mathcal X$. This intensity is the normalized derivative of \eqref{endALL}. 

The conditional intensity function at time $K$ derived from \eqref{endALL} is conceptually very   different  from the intensity function that is  output by a sequence of $K$ one-step intensity filters (i.e., the filters discussed in Section \ref{iFilter}) that  close the one-step Bayesian recursions by   approximating the one-step intensities. Making  $K$ such approximations can result in artifacts that accumulate over time.  In contrast,  a sliding batch  implementation of a length $K$  interval filter makes only one approximation. This   reduces the risk of artifacts at the cost of  increased computational complexity. Sliding batch intensity filters remain to be studied.   

\subsection{PGFs for  the Number of Trajectories}
Denote the number of trajectories that are present at time $k$ by $N_k$, and let  $N_{1:K}\equiv (N_1,\ldots,N_K)$.  

\subsubsection{Prior Distribution}
Evaluating the joint PGFL \eqref{smootherer}  for $g_k(\cdot)=1$ for all $k$ gives  the PGFL of the prior distribution of all  target trajectories:   
$$
\Psi^{\mathrm{Batch}}(h_{1:K})\equiv\left.\Psi^{\mathrm{Batch}}(h_{1:K},g_{1:K})\right|_{\substack{g_k(\cdot)\,=\,1\\k=1,\ldots,K}}.
$$
Setting  ${h_k(x_k)\,=\,z_k}$ for all $k$ gives the PGF of the prior distribution of the number $N_{1:K}$ of trajectories,
\begin{align}
    \label{PriorNumbers}
\mathrm G^{N_{1:K}}(z_{1:K})\equiv\left.\Psi^{\mathrm{Batch}}(h_{1:K})\right|_{\substack{h_k(x_k)\,=\,z_k\\ k=1,\ldots,K}}.
\end{align}
The prior probability ${\Pr\{N_1=n_1,\ldots,N_k=n_K\}}$ is the coefficient of $z_1^{n_1} \cdots z_K^{n_K}$ in the  series expansion of \eqref{PriorNumbers} about the origin. 

Consider the special case when the branching processes are independent of target state, i.e., when \eqref{targetNo} takes the form 
\begin{align}
    \label{targetNoINDEP}
    \mathrm G^{B_k}(z)\equiv \mathrm G^{B_k|x_{k-1}}(z)=\textstyle\sum_{b=0}^\infty \Pr\{B_k=b
    \}\, z^{b}.
\end{align}
Unrolling the recursion gives  the joint PGF 
\begin{equation}
\label{GalWat}
\begin{split}
    \begin{aligned}
    &\mathrm G^{N_{1:K}}\big(z_{1:K}\big)\\
    &\,\,\,=\mathrm G^{B_1}\big(z_1 \mathrm G^{B_2}\big(z_2  \cdots \mathrm G^{B_{K-1}}\big(z_{K-1} \mathrm G^{B_K}(z_K)\big)\cdots\big)\big).
\end{aligned}
\end{split}
\end{equation}
The form of this PGF proves that $N_{1:K}$ is a time inhomogeneous branching process (see \cite[Ch. I.13.1]{1963Harris} and the references therein).
Setting  ${z_1=\cdots=z_{K-1}=1}$ gives the PGF of  the number of trajectories at the final time $K$,   
\begin{equation}
\label{GalWat2}
\begin{split}
    \begin{aligned}
    \mathrm G^{N_{K}}\big(z_{K}\big)
    =\mathrm G^{B_1}\big(\mathrm G^{B_2}\big(  \cdots \mathrm G^{B_{K-1}}\big( \mathrm G^{B_K}(z_K)\big)\cdots\big)\big).
\end{aligned}
\end{split}
\end{equation}
Setting ${z_2=\cdots=z_{K}=1}$ gives the PGF of the number of trajectories at time ${k=1}$, namely, $\mathrm G^{B_1}\big(z_1\big)$. 
If  the branching process is time homogeneous, i.e., if $\mathrm G^{B_k}(\cdot)\equiv \mathrm G(\cdot)$ for all $k$, then $\mathrm G^{N_{K}}\big(z_{}\big)
    =\mathrm G^{\mathrm{o}(K)}(z)$, 
where $\mathrm G^{\mathrm{o}(K)}$  denotes  the repeated composition of the PGF $\mathrm G(z)$ with itself $K$ times.

\subsubsection{Posterior Distribution} 
Evaluating the joint PGFL \eqref{smootherer}  for $h_k(x_k)=z_k, x_k\in\mathcal X_k,$ for all $k$ gives the joint PGF-PGFL of $N_{1:K}$ and all  measurement sequences,  
\begin{align}
    \label{Numbers}
    \Psi^{N_{1:K}}(z_{1:K},g_{1:K})\equiv\left.\Psi^{\mathrm{Batch}}(h_{1:K},g_{1:K})\right|_{\substack{h_k(x_k)\,=\,z_k\\ k=1,\ldots,K}}. 
\end{align}
It is computed via the backward recursion.   Denote  the measurement set at time $k$ by  $\mathbf y_{1:m_k}$, where for simplicity we assume $m_k\geqslant 1$.  Let $\mathbb Y_K=(\mathbf y_{1:m_1},\ldots, \mathbf y_{1:M_K})$.  For $m\geqslant 1$, let  $\mathbf 0_{m}=(0,\ldots,0)\in\mathbb R^m$ and  $\mathbf 1_{m}=(1,\ldots,1)\in\mathbb R^m$.  Using an obvious extension of the notation in Eqn. \eqref{BayesIsBeautiful} with ${\boldzero_K=({\bf 0}_{m_1},\ldots,{\bf 0}_{m_K})}$ and  ${\boldone_K\equiv (\mathbf 1_{m_1},\ldots,\mathbf 1_{m_K})}$, the  PGF of $N_{1:K}$ conditioned on $\mathbb Y_K$ is the normalized derivative of the secular function of \eqref{Numbers} for  the measurement sets in $\mathbb Y_K$, 
\begin{align}
\label{numbers}
\mathrm G^{N_{1:K}}(z_{1:K}|\mathbb Y_K)=\frac{\mathrm D^{(\boldzero_K,\boldone_K)}\Psi^{N_{1:K}}(z_{1:K},\boldzero_K))}{\mathrm D^{(\boldzero_K,\boldone_K)}\Psi^{N_{1:K}}(\boldone_K,\boldzero_K))}.
\end{align}
The coefficient of the monomial $z_1^{n_1} \cdots z_K^{n_K}$ in the power series expansion of \eqref{numbers} about the origin is the conditional probability ${\Pr\{N_1=n_1,\ldots,N_k=n_K|{\mathbb Y_K\}}}$.


\bigskip
{\color{black}
\section{Assignment Problems}
\label{GFforAssignments}}

Assignments of sensor measurements to targets are  events whose probabilities are derived from the  PGFL. The technique is general, but we  eschew the general case and choose instead to illustrate it  by example.  The technique  was encountered earlier in the single target IMM filter (Section \ref{IMM Filters}) and its  multitarget generalization (Section \ref{JointIMMfilters}).  Several additional examples are given in this section.  

We use the secular method \cite{2014Streit} to derive the assignment probabilities.  The secular method circumvents entirely the need to calculate probabilities using  functionals and functional derivatives. Using this  technique, the derivation of Bayesian filters involves only  ordinary functions and  their derivatives.   It is reviewed  in the Appendix (Eqns.  \eqref{deltatrain}-\eqref{connection} and Eqns. \eqref{easyBi}-\eqref{BayesSec}).


\smallskip
\subsection{Example 1:  Single Target PDA}
\label{labeledPDAassign}
The PGFL of the PDA filter with a general false alarm process is given by  \eqref{PDA}. The  PGFL conditional on  $x\in\mathcal X$  is 
\begin{align}
    \label{assignPDA}
&\Psi^{\mathrm{PDA}}(g|x)\\
&=\mathrm G^{\mathrm{FA}}\!\left(\textstyle\int_{\mathcal Y}g(y) p^{\mathrm{FA}}(y) \,\mathrm dy\right) \!\left(Qd(x)\!+\!P\!d(x)\!\textstyle\int_{\mathcal Y}  g(y) p(y|x)\,\mathrm dy\right)\!.\nonumber
\end{align}
To see this, note that the conditioning here is the reverse of  the conditioning in \eqref{BayesIsBeautiful}. Adapting the method  in the Appendix to this problem gives \eqref{assignPDA}. Further details are omitted.  

Let    ${{\bf y}= (y_1,\ldots,y_{M})},\,{M\ge 1}$, denote the sensor measurement set.  The 
measurement $y_m$ is labeled by  $j_{m}^0$ if it is a false alarm and by $j_{m}^1$ if it is generated by the target.  The measurement labels  $j_{m}^0$ and $j_{m}^1$ are  complex-valued variables, but they are  {\it{not}} indeterminates of random variables. 
Let ${J^0=\{j_{1}^0,\ldots,j_{M}^0\}}$,    ${J^1=\{j_{1}^1,\ldots,j_{M}^1\}}$, and ${J=\{J^0, J^1\}}$.   Let  ${\gamma =(\gamma_1,\!\ldots,\gamma_M)\in\mathbb C^M}$.  Define   labeled delta trains for the false alarms  and  target measurements by, resp.,  \begin{subequations}\label{subsecab}
\begin{align}
    \label{subSec1}g_{\mathrm{sec}}^0(y)&=\textstyle\sum_{m=1}^{M} \gamma_{m} \,\delta_{y_{m}}(y)\, j_{m}^0 
    \\
    \label{subSec2}g_{\mathrm{sec}}^1(y)&=\textstyle\sum_{m=1}^{M} \gamma_{m}\,\delta_{y_{m}}(y) \, j_{m}^1 .
\end{align}
\end{subequations}
Substituting  \eqref{subSec1}  and  \eqref{subSec2} into \eqref{assignPDA}  gives the the secular function 
\begin{equation}
\label{PDAexplicitSEC}
\begin{split}
    &\Psi_{\mathrm{sec}}^{\mathrm{PDA}}(\gamma;x,J)\\
    &\,\,\,= \mathrm G^{\mathrm 
    {FA}}\!\left(\textstyle\sum_{m=1}^{M} \gamma_{m} \,a_m \,j_{m}^0 \right )\!\left(b_0 +\textstyle\sum_{m=1}^{M}  \gamma_{m}  \,b_m\, j_{m}^1\right),
\end{split}
\end{equation}
where the coefficients $a_m$ and $b_m$ are 
\begin{align}
\label{cherryOnTop}
a_m&= p^{\mathrm{FA}}(y_{m}),\,\,\, b_{0}= Qd(x), \,\,\, 
    b_{m}= \,P\!d(x) \,p(y_{m}|x).    
\end{align}
The expression \eqref{PDAexplicitSEC} is identical to \eqref{mumboPDA} for ${j_m^0=j_m^1=1}$.  
The  mixed first-order derivative  of $\eqref{PDAexplicitSEC}$ with respect to  $\gamma_{}$ is    
\begin{align}
\label{tripleJunk}
    &{\mathrm D}^{({\mathbf 1}_M)} \Psi_{\mathrm{sec}}^{\mathrm{PDA}}({\mathbf 0}_M;x,J)\nonumber\\
    &\quad = A_0(\mathbf y|x)\!\left( \prod_{m=1}^M j_{m}^0\right)  
    \,+\sum_{m=1}^M A_m(\mathbf y|x)\!\!\left(j_m^1\!\prod_{m'\ne m}^M j_{m'}^0 \right),
\end{align}
where the  coefficients  are (compare to \eqref{dullville})   
\begin{align*}
    A_0(\mathbf y|x)&\equiv M!\,\Pr\{M\}\, b_0\,\textstyle  \prod_{m=1}^M a_m
    \\
A_m(\mathbf y|x)&\equiv (M\!-\!1)!\,\Pr\{M\!-\!1\} \,b_m\,\textstyle\prod_{m'\ne m}^M
    a_{m'}.
\end{align*}
The products in big parentheses are  monomials in the measurement labels  $J$.  The monomials  encode the $M+1$ feasible measurement to target assignments (including, tediously, the assignments of measurements to false alarms). 
The assignment probabilities are found by normalizing \eqref{tripleJunk}.  The normalizing constant is  $\sum_{m=0}^M A_m(\mathbf y|x)$.  Reverting to the original variables and simplifying gives     
\begin{align}
    \Psi(J|x,{\bf y})=\beta_0\prod_{m=1}^M j_{m}^0 + \sum_{m=1}^M \beta_m \left(j_m^1\prod_{m'\ne m}^M j_{m'}^0 \right)\!,\label{PDAassignPosterior}
\end{align}
where  
$$\beta_0=\frac{Qd(x)}{Qd(x)+\frac{\Pr\{M-1\}}{M\Pr\{M\}}\sum_{m=1}^M \frac{P\!d(x) p(y_{m}|x)}{p^{\mathrm{FA}}(y_{m})}}$$
is the conditional probability that the target in state $x$ is not detected  and all  measurements are labeled as false alarms, and  
$$\beta_m=\frac{\frac{\Pr\{M-1\}}{M\Pr\{M\}}\frac{P\!d(x) p(y_{m}|x)}{ p^{\mathrm{FA}}(y_{m})}}{Qd(x)+\frac{\Pr\{M-1\}}{M\Pr\{M\}}\sum_{m=1}^M \frac{P\!d(x) p(y_{m}|x)}{ p^{\mathrm{FA}}(y_{m})}}$$
is the conditional probability that  $y_m$ is labeled by $j_m^1$, i.e., is generated by the target  in state $x$, and the other measurements are labeled as false alarms. 

Setting $j^0_m=1$ for all $m$ gives the simplified list of  measurement-to-target  assignment probabilities, 
\begin{align*}
    \Psi(J_1|x,{\bf y})&=\beta_0 + \textstyle \sum_{m=1}^M \beta_m \,\,j_m^1.\label{PDAassignPosteriorJ1only}
\end{align*}
We included the false alarm  labels $j^0_1,\ldots,j^0_M$   in this example to illustrate that they are unnecessary.  The only labels we  need  are those that indicate which measurement is assigned to the target.   

\subsection{Example 2: Multitarget J-PDA }
\label{labeledJPDA}
The assignment problem can be studied  starting with  the marginal PGFL \eqref{margieJPDA}. This  form of the assignment problem  is not considered  here, but it can be studied with the  methods  in this section. 

We start with the PGFL (cf. \eqref{gfJPDA})  of the J-PDA filter for a Poisson false alarm process and $N$ independent BMD targets, namely,   
\begin{align}
\nonumber
    \begin{split}
        &\Psi^{\mathrm{J\mhyphen PDA}}(h,g)=\exp\left(-\lambda^{\mathrm{FA}}+\lambda^{\mathrm{FA}}\textstyle\int_{\mathcal Y}\,g(y) p^{\mathrm{FA}}(y)\,\mathrm dy \right )\\
    &\,\,\times\!\prod_{n=1}^N  \!\int_{\!\mathcal X_n}\!\!\! h_n(x) p_n^{\!-}\!(x)\! \left(Qd_n(x)\!+\!P\!d_n(x)\!\textstyle\int_{\mathcal Y}  g(y) p_n(y|x_n)\,\mathrm dy\right)\mathrm dx,
    \end{split}
\end{align} 
where the $n^{\mathrm{th}}$ term in the product is the PGFL for  target $n$ (cf.  \eqref{BasicBMD}). Let  ${{\bf x}\equiv (x_1,\ldots,x_N)}$, where  ${x_n\in\mathcal X_n}$ is the state of target $n$.    The same method used to derive \eqref{assignPDA} gives the PGFL of the measurements conditioned on $\mathbf x$,  
\begin{equation}
\label{JPDAconditioanlonXn}
\begin{aligned}
    \Psi^{\mathrm{J\mhyphen PDA}}&(g|\mathbf x)=\exp\left(-\lambda^{\mathrm{FA}}+\lambda^{\mathrm{FA}}\textstyle\int_{\mathcal Y}\,g(y) p^{\mathrm{FA}}(y)\,\mathrm dy \right )\\
    &\times\prod_{n=1}^N  \left(Qd_n(x)+P\!d_n(x)\textstyle\int_{\mathcal Y}  g(y) p_n(y|x_n)\,\mathrm dy\right). 
\end{aligned} 
\end{equation}
As in the previous subsection, let    ${{\bf y}= (y_1,\ldots,y_{M})},\,{M\ge 1}$.  The assignment of measurement ${y_m\in \mathbf y}$  to target $n$ in state  $x_n$ is given the complex-valued label $j^n_m$,   ${1\leqslant m\leqslant M}$. As shown in the previous subsection, we do not need to label  assignments to false alarms.  Let $J^n_M=\{j^n_1,\ldots,j^n_M\}$ and $J=\{J^1_M,\ldots,J^N_M\}$.    The labeled delta train for the assignments of measurements  to  target $n$ is 
\begin{align}
    g_{\mathrm{sec}}^n(y)=\textstyle\sum_{m=1}^{M} \gamma_{m} \, \delta_{y_{m}}(y)\,j_{m}^n .
\end{align}
Substituting into \eqref{JPDAconditioanlonXn} gives the  secular function, 
\begin{equation}
\label{J-PDAexplicitSEC}
\begin{split}
    &\Psi_{\mathrm{sec}}^{\mathrm{J\mhyphen PDA}}(\gamma;\mathbf x,J)\equiv  \,\exp\!\left(a_0 +\textstyle\sum_{m=1}^{M} \gamma_{m} \,a_m(y_m)  \right )
    \\&\qquad\times\prod_{n=1}^N\left(c^{n}_0(x_{n}) +\textstyle\sum_{m=1}^{M}  \gamma_{m}  \,c^{n}_m(y_m,x_{n})\,j_m^n\right),
\end{split}
\end{equation}
where  
\begin{align*}
    a_0&=-\lambda^{\mathrm{FA}},\qquad  \quad \,\,\,\,a_m(y_m)=\lambda^{\mathrm{FA}} p^{\mathrm{FA}}(y_m)\\
    c_0^n(x_n)&=Qd_n(x_n),\quad
    c_m^n(y_m,x_n)=P\!d_n(x) \,p_n(y_{m}|x_n).
\end{align*} 
Normalizing gives 
\begin{align}
\label{NPstartshere}
    \Psi(J|{\bf x},{\bf y})=\frac{{\mathrm D}^{({\mathbf 1}_M)} \Psi_{\mathrm{sec}}^{\mathrm{J\mhyphen PDA}}({\mathbf 0}_M;\mathbf x, J)}{{\mathrm D}^{({\mathbf 1}_M)} \Psi_{\mathrm{sec}}^{\mathrm{J\mhyphen PDA}}({\mathbf 0}_M;\mathbf x,J=\mathbf 1)}. 
\end{align}
The derivative in \eqref{NPstartshere} is a multivariate polynomial in the variables in $J$.  The derivative is also  an NP-hard calculation. Nonetheless, once  it is  calculated, the coefficients of the monomials are  the conditional probabilities  of the feasible assignments of  measurements in $\mathbf y$ to the $N$ targets when they are in state $\mathbf x$.  

To see that the derivative is NP-hard, it suffices to consider the special case of no false alarms and no missed detections, so that ${N=M}$. We also limit our interest to assignments of measurements in $\mathbf y$ to target $n$, so we let $j_m^{n'}=1,\,n'\ne n$.  In this  case  the secular function \eqref{J-PDAexplicitSEC} is 
\begin{equation}
\label{J-PDAexplicitSECreduced}
\begin{split}
    \widetilde\Psi_{\mathrm{sec}}^{\mathrm{J\mhyphen PDA}}(\gamma;J_M^n)&\equiv \left(\sum_{m=1}^{M}  \gamma_{m}  \,c^{n}_m(y_m,x_{n})\,j_m^n\right)\\
    &\times \prod_{n'=1,n'\ne n}^M\!\!\left(\sum_{m=1}^{M}  \gamma_{m}  \,c^{n}_m(y_m,x_{n})\right).
\end{split}
\end{equation}
Recall the generic derivative\footnote{To derive    \eqref{genericDeriv}, expand the product into the $L$-fold  multiple  sum, so that ${\textstyle{\prod_{k=1}^L \!\left(\sum_{\ell=1}^L a_{k\ell}\gamma_\ell\right)\!=\!\sum_{\ell_1\!=1}^L\cdots\sum_{\ell_L\!=1}^L a_{1,\ell_1}\!\cdots a_{L,\ell_L}\gamma_{\ell_1}\!\cdots \gamma_{\ell_L}}}$, 
and  take the derivative term-by-term using the multivariate  Cauchy integral  \eqref{cauchy}.}  
\begin{align}
\label{genericDeriv}
    \left.\frac{\partial^L}{\partial \gamma_1\cdots\partial \gamma_L}\right|_{\gamma=0}\prod_{k=1}^L\left(\sum_{\ell=1}^L a_{k\ell}\gamma_\ell\right)=\mathrm{perm}(A),
\end{align}
where $A=\big[a_{k\ell}\big]\in\mathbb C^{L\times L}$ is a square  matrix,  
\begin{align}
\label{permanentMatrix}
\mathrm{perm}(A)=\textstyle\sum_{\sigma\in \mathrm{Sym}(L)} \,a_{1,\sigma(1)} \,a_{2,\sigma(2)}\cdots a_{L,\sigma(L)}
\end{align}
is the matrix permanent, and $\mathrm{Sym}(L)$ is the set of $L!$ permutations of the first $L$ positive integers. The fastest known algorithm for computing $\mathrm{perm}(A)$ requires   $O(2^L L^2)$ arithmetic operations. 

Using  \eqref{genericDeriv}, the derivative 
$$
{\mathrm D}^{({\mathbf 1}_M)} \widetilde\Psi_{\mathrm{sec}}^{\mathrm{J\mhyphen PDA}}({\mathbf 0}_M;\mathbf x,J_M^n=\mathbf 1_M)=\mathrm{perm}(\Lambda),
$$
where 
$\Lambda=[\lambda_{mn}]\in\mathbb R^{M\times M}$ and $\lambda_{mn}=c_m^n(y_m,x_n)$. The derivative 
$$
{\mathrm D}^{({\mathbf 1}_M)} \widetilde\Psi_{\mathrm{sec}}^{\mathrm{J\mhyphen PDA}}({\mathbf 0}_M;\mathbf x, J_M^n)
$$
is also evaluated using \eqref{genericDeriv} but retaining the measurement assignment  labels $J^n_M=\{j^n_1,\ldots,j^n_M\}$. It is straightforward to show from the definition \eqref{permanentMatrix} (details omitted) that the derivative  is 
$$
{\mathrm D}^{({\mathbf 1}_M)} \Psi_{\mathrm{sec}}^{\mathrm{J\mhyphen PDA}}({\mathbf 0}_M;\mathbf x, J_M^n)=\textstyle\sum_{m=1}^M \lambda_{mn}\,{\mathrm{perm}(\Lambda_{mn})}\,j_m^n,
$$
where   $\Lambda_{mn}$ is the  ${(M\!-\!1)\times (M\!-\!1)}$ matrix obtained from $\Lambda$ by deleting row $n$ and column $m$. Thus, the expression \eqref{NPstartshere} evaluated for the secular function \eqref{J-PDAexplicitSECreduced} is   
 \begin{align}
 \label{AUCW}
     p(J^n_M|{\bf x},{\bf y})=\sum_{m=1}^M \left(\lambda_{mn}\frac{\mathrm{perm}(\Lambda_{mn})}{\mathrm{perm}(\Lambda)}\right)j^n_m\,.
 \end{align}
The coefficient of $j^n_m$ is the posterior probability that measurement $m$ is assigned to target $n$ in the special case of no missed  detections and no false alarms. 

Expressions for more general  problems can be derived from \eqref{NPstartshere} by extending the definition of a matrix permanent from square to rectangular matrices  \cite{1972Achkasov}\cite{2023ORourke}\cite{2004Uhlmann}\cite{1995ZhouBose}\cite{2017CrouseWillett}.  These results are not presented here.  

\subsection{Example 3: Multiple Sensor J-BMD}
\label{labeledMS}
Assignment probabilities are used  for  track-to-track fusion across multiple sensors (see \cite{Ferry2006} and the references therein).  The modeling assumptions are those of the multisensor version of the labeled \mbox{J-BMD} problem of Section \ref{MultiSensorBMD}, together with the assumption that no false tracks are generated by any of the sensor-level tracking systems.  The relevant  PGFL is   \eqref{GFBM1NmultiEll}. (If false tracks are present, the PGFL     \eqref{GFBM1NmultiEll}  is multiplied by the PGFLs of the  false  sensor-level tracks.)  

Let ${\bf x}\equiv(x_1,\ldots,x_N)$, where ${x_n\in  \mathcal X_n}$. Substitute the weighted deltas $h_n(x)=\alpha_n \delta_{x_n}(x),\, x\in\mathcal X_n$. The delta  operators $h_{1:N}(\cdot)$   evaluate the integrands at ${x_{1:N}}$. The result is a function of $\alpha_{1:N}$.  The first-order mixed derivative with respect to $\alpha_{1:N}$ evaluated  at ${\bf 0}_{1:N}$ is the secular PGFL. Using  \eqref{bmdofnSensorEll} of   $\Psi^{\mathrm{BMD}}_{\ell|n}(g_\ell|x_n)$,  the conditional PGFL is   
\begin{align}
\label{GFBM1NmultiEllrrr}
\Psi^{\mathrm{MS\mhyphen J\mhyphen BMD}}(g_{1:L})&=\Psi^{\mathrm{MS\mhyphen J\mhyphen BMD}}(g_{1:L}|\bf x)\nonumber\\
&\equiv  \prod_{n=1}^N \prod_{\ell=1}^L \Psi^{\mathrm{BMD}|x_n}_{\ell|n}(g_{\ell}). 
\end{align}
Let ${{\bf y}\equiv ({{\bf y}}^1,\ldots,{{\bf y}}^L)}$, where ${{\bf y}^\ell= (y_1^\ell,\ldots,y_{M_\ell}^\ell)},\,{M_\ell\geqslant 1}$, is the set of measurements from sensor $\ell$. Let $M=\sum_{\ell}M_\ell$.  (We omit the cases for which  one or more sensors have no measurements; the details are straightforward, but tedious.)   The subscripted-superscripted  complex-valued indeterminate variable $j^{n}_{\ell,m}$ corresponds to the assignment of  sensor measurement ${y_m^\ell}$, ${1\leqslant m\leqslant M_\ell,\,1\leqslant \ell\leqslant L}$, to  target $n$ in  state ${x_n,\, 1\leqslant n\leqslant N}$.   The set of all ${N\!M\!L}$ indeterminates is denoted by $J$.   
Let ${\gamma^\ell=(\gamma_1^\ell,\ldots,\gamma_{M_\ell}^\ell)}$ and ${\gamma =(\gamma^1,\ldots,\gamma^L)}$.  Substitute the labeled delta trains 
\begin{align}
    g_\ell(y)=\sum_{m_\ell=1}^{M_\ell} \gamma_{m_\ell}^\ell \, j^{n}_{\ell,m_\ell} \,\delta_{y^\ell_{m_\ell}}(y),\quad y\in\mathcal Y_\ell,
\end{align}
into  \eqref{GFBM1NmultiEllrrr}. Eliminating the  integrals  using  the delta operators and  interchanging the products over $n$ and $\ell$ gives  the secular function 
\begin{align}
\label{GFBM1NmultiEllrrrrrr}
\Psi&^{\mathrm{MS\mhyphen J\mhyphen BMD}}(\gamma)
\equiv  \prod_{\ell=1}^L \Phi_\ell(\gamma^\ell) ,
\end{align}
where, for sensor $\ell$,  
\begin{align}
\label{sensorLsecular}
\Phi_\ell(\gamma^\ell) = \prod_{n=1}^N\Bigg(b_n^\ell(x_n)+\sum_{m_\ell=1}^{M_\ell}\gamma_{m_\ell}^\ell  \,c_{n}^{\ell,m_\ell}(x_n)\,j^{n}_{\ell,m_\ell}\Bigg)
\end{align}
and 
\begin{align}
    b_n^\ell(x_n)&= Qd_n^\ell(x_n)\nonumber\\
c_{n}^{\ell,m_\ell}(x_n)&=P\!d_n^\ell(x_n) \,p_n^\ell(y_{m_\ell}^\ell|x_n).\nonumber
\end{align}
The mixed first-order derivative of \eqref{GFBM1NmultiEllrrrrrr}   evaluated at $\gamma=0$  is the joint pdf of $({\bf x},{\bf y})$ if $J=\mathbf 1$, i.e., all  indeterminate variables  $j^{n}_{\ell,m}=1$. We  retain the  variables $J$, but still take the mixed derivative, so that 
\begin{equation}
     \label{joinglabelGF}
\begin{aligned}
    p({\bf x},{\bf y};J)&=\left.\mathrm  D^{(\gamma^1\!,\ldots,\gamma^L)}\right |_{\gamma^1=\cdots =\gamma^L=0} \Psi^{\mathrm{MS\mhyphen J\mhyphen BMD}}(\gamma)\\
    &=\prod_{\ell=1}^L \left\{\left . \mathrm D^{(\gamma^\ell)} \right |_{\gamma^\ell=0}\Phi_\ell(\gamma^\ell)\right\},
\end{aligned}
\end{equation}
where we used the fact that  $\Phi_\ell(\gamma^\ell)$ depends  on $\gamma^\ell$ and  not on the variables $\{\gamma^{\ell'}\!,\,\ell'\ne \ell\}$. 

The derivative of $\Phi_\ell(\gamma^\ell)$ is straightforward but messy to write in general, so here we give only the special case $M_\ell=2$. There are no false alarms, so  $M_\ell\leqslant N$. For illustrative purposes, we consider the case  $M_\ell<N$.   Then: 
\begin{equation}
\label{nastyexpression}
\begin{split}
    &\!\!\!\frac{\partial^2 }{\partial \gamma_1^\ell \,\partial \gamma_2^\ell} \bigg |_{\gamma_1^\ell=\gamma_2^\ell=0}\Phi_\ell\big(\gamma^\ell\big)\\
    &=\sum_{n\!'=1}^N\!\sum_{\substack{\,n\!''\!=1\\\,n\!''\ne n\!'}}^N \!\!\left\{\!\!\left(\!\prod_{\substack{n=1\\ \,\,n\ne n\!',n\!''}}^N \!\!b_n^\ell(x_n)\right)\!c_{n'}^{\ell,1}(x_{n'})\, c_{n''}^{\ell,2}(x_{n''}) \right\}j^{n'}_{\ell,1} \,j^{n''}_{\ell,2}.
\end{split}
\end{equation}
The term in braces is the probability that (in sensor $\ell$) measurement 1  is assigned  to target $n'$, that  measurement 2 is assigned to target $n''$, and that no other targets are  detected. This term is the coefficient of the monomial  $j^{n'}_{\ell,1} \,j^{n''}_{\ell,2}$.   The double sum is the sum over all feasible assignments of two of  the $N$ targets to the two measurements in sensor $\ell$. 

The joint distribution \eqref{joinglabelGF} is a product over the $L$ sensor level derivatives, each of which is a sum of the $N\!M$ monomials in   ${J^\ell\equiv \{j^{n}_{\ell,m}:1\leqslant m\leqslant  M_\ell,\,1\leqslant n\leqslant  N\}}$, as illustrated by \eqref{nastyexpression}. Expanding the product yields a sum of the  $N\!M\!L$  monomials in $J=\{J^1,\ldots,J^L\}$.  Each of these  monomials identifies a feasible assignment, and its coefficient is the probability of the assignment.

Despite appearances,   \eqref{joinglabelGF} is {\it{not}} the joint PGF of the set of feasible multisensor assignments because {\it{assignments are not  random variables.}}   
They do not appear in the joint PGFL \eqref{GFBM1NmultiEllrrr}. If we wish, we can define   {\it{de novo}} random variables whose probability law is defined by the PGF  
\begin{align}
\label{MoT1}
    \mathrm G^{\mathrm{MtoT}}(J|{\mathbf x},{\mathbf y})\equiv \frac{ p({\bf x},{\bf y};J)}{ p({\bf x},{\bf y};J=\mathbf 1_{N\!M\!L}) },
\end{align}
where $\mathbf 1_{N\!M\!L}$ is an array of size $N\!\times \!M\!\times\! L$ all of whose entries are equal to 1. 
The product  \eqref{joinglabelGF} implies that the {\it{de novo}}  assignment variables are   conditionally independent from sensor to sensor.    


\bigskip
{\color{black}
\section{Saddle Point Approximation}
\label{saddlept}
}

The exact particle weights  are high computational complexity calculations in many  particle filters for tracking applications. We show that the saddle point method  can be used to approximate  particle weights  for nearly all the filters in this paper.  The  method  is widely used in quantum mechanics and  has a long history in applied mathematics and statistics.

The saddle point approximation is an  easily understood numerical procedure. The derivation, however, involves mathematical methods rarely needed in tracking problems.  Readers interested only in the algorithm can skip directly to the numerical problem, \eqref{argmnin}.  The approximation is based on three key steps:
\begin{itemize} 
\item Particle weights are identically equal  to derivatives of the PGFL. 
\item Derivatives are identical to multivariate Cauchy integrals. 
\item Integrals are approximated at their saddle points.  
\end{itemize}
The first  step is analytical.    The second step is crucial---it is the segue from analysis to numerics (and asymptotic expansions \cite{FS}).   The third step is strictly numerical \cite{Jensen}.  The derivation   involves complex variables,  but  only real-valued numerical methods are needed to compute the  particle weight approximation.    

To review (see, e.g., \cite{BTKF}), SIR particle filters start with a ``large'' number of equally weighted i.i.d. samples of the posterior at the previous scan time.  Using the target motion model, these particles are used to generate particles that sample the predicted target state at the current scan time \eqref{SIR}. Each  particle is weighted by the likelihood of the current scan data conditioned on the particle. This is a high complexity calculation in many filters and  cannot be avoided---the  weights are used to resample the particles.  Simply put, SIR particle filters fail without resampling.     

Enter the saddle point method.  

We illustrate the saddle point method by applying it to the  labeled  J-PDA filter \cite{streit2015},\cite{FSS}. The  joint PGFL for J-PDA with $N$ targets is  \eqref{gfJPDA}.  J-PDA has high computational complexity; more precisely, the exact filter is known to be NP-hard. 

Given the scan measurement set  ${\mathbf y=\{y_1,\ldots,y_m\}\subset\mathcal Y}$ and a particle $x\in\mathcal X_n$  for target $n$, we compute the particle filter weight $w(x)\equiv p_{\mathrm{J\mhyphen PDA}} (\mathbf y|x)$ using the marginal PGFL \eqref{margieJPDA} for target $n$. This is the reverse of the conditioning in \eqref{BayesIsBeautiful}.  The marginal \eqref{margieJPDA}  is linear in $h_n$, so the PGFL of the  particle weight is, with  Poisson false alarms,  
\begin{equation}
    \begin{split}
    \label{BayesGFLPF}
    &\Psi(g|x)=\exp\left(-\lambda^{\mathrm{FA}}+\lambda^{\mathrm{FA}}\textstyle\int_{\mathcal Y}\, g(y) p^{\mathrm{FA}}(y) \,\mathrm dy \right )\\
    &\qquad \times \! \left(  Qd_n(x)+P\!d_n(x)\textstyle\int_{\mathcal Y} \, g(y) p_n(y|x)\,\mathrm dy\right)\\
    &\!\!\times \!\!\prod_{n'\ne n}\!  \int_{\!\mathcal X_{n'}}\!\!  p_{n'}^{\!-}(x)\! \left(Qd_{n'}(x)+P\!d_{n'}(x)\textstyle\int_{\mathcal Y}  g(y) p_{n'}(y|x)\,\mathrm dy\right)\mathrm dx.
    \end{split}
\end{equation}
Note that the prior $p_{n}^{\!-}(x)$ of target $n$ is absent from \eqref{BayesGFLPF} because it cancels out in the ratio that defines the Bayes PGFL. (As a check, readers may verify that $\Psi(g|x)=1$ for $g(\cdot)=1$.)   Substituting $g(y)=\sum_{k=1}^m g_k \delta_{y_k}(y)$ into \eqref{BayesGFLPF} gives the  secular function   
\begin{equation}
    \label{detailsdetals}
\begin{split}&\Psi(g_{1:m}|x)=\exp\left(-\lambda^{\mathrm{FA}}+\lambda^{\mathrm{FA}}\textstyle\sum_{k=1}^m\, p^{\mathrm{FA}}(y_k) \,g_k \right )\\
    &\times\! \Big(b_n(x)+\sum_{k=1}^m c_{nk}(x,y_k)\, g_k\Big)\! \prod_{n'\ne n}\!\Big(\bar b_{n'}+\sum_{k=1}^m \bar c_{n'k}(y_k) \,g_k\Big),
\end{split}
\end{equation}
where, for all $1\le n\le N$, we define the convenient notation 
\begin{equation*}
\begin{array}{l|l}
    b_n(x)= Qd_n(x) & c_{nk}(x,y_k)=P\!d_n(x) p_n(y_k|x)
   \\
     \bar b_n=\textstyle\int_{\mathcal X_n} p_n^-(x) b_n(x)\,\mathrm dx  & 
     \bar c_{nk}(y_k)=\textstyle\int_{\mathcal X_n} p_n^-(x) c_{nk}(x,y_k)\,\mathrm dx .
    \end{array}
\end{equation*}
When the distributions are linear-Gaussian, the integrals can be evaluated explicitly; otherwise, they are evaluated numerically, e.g., by using the current target  particles.   

The  weight $w(x)$ of a particle for target $n$  at $x\in\mathcal X_n$ is  the mixed first-order derivative of  $\Psi(g_{1:m}|x)$, 
\begin{align}
    \label{weight}
    w(x)=\left.\frac{\partial ^m}{\partial g_1\cdots\,\partial g_m}\right|_{g_1=\,\cdots\,=g_m=\,0}\Psi(g_{1:m}|x).
\end{align}
This expression is exact.  The complex-valued multivariate function $\Psi(g_{1:m}|x)$ is  analytic in each variable separately, so it is  jointly analytic in all $m$ variables. Consequently, the derivative is identical to the  multivariate Cauchy integral 
\begin{align}
    \label{cauchy}
    w(x)=\frac{1}{(2\pi i)^m}\oint_{C(r_1)}\!\!\cdots\,\oint_{C(r_m)} \frac{\Psi(g_{1:m}|x)}{g_1 \cdots g_m}\,\frac{\mathrm dg_1\cdots\,\mathrm dg_m}{g_1 \cdots g_m},
\end{align}
where ${C(r_k)}$ is a  circle centered at the origin of the $g_k$-complex plane of radius ${r_k > 0}$. 

We use the saddle point method to approximate the multivariate  integral. The method is carefully discussed  in many places, e.g. \cite{FS} and \cite{Jensen}.  We do not review it here, but instead go straight to the numerical method. 

Let $r=(r_1,\ldots,r_m)$, where $r_k=\mathrm{Re}{\,g_k}$.  To find the radii of the circles, define  ${\phi:\mathbb R^m_{+}\rightarrow \mathbb R_+}$ by
\begin{align}
\label{logPsi}
    \phi(r_{}) =\log \Psi(r_{}|x)-\sum_{k=1}^m \log r_k,\,\,\,\, r_k>0,\,\,\,1 \leqslant k\leqslant m.
\end{align}
Since $e^{\phi(r)}$  is real-valued and non-negative for $r\in\mathbb R^m_{+}$, we  use   numerical methods to solve  the (multivariate) minimization problem 
\begin{align}
    \label{argmnin}
\boxed{
\thickhat r\, =\argmin_{r\,> \,0}\,\phi(r_{}).
}
\end{align}
The point $\thickhat r= (\thickhat r_1,\ldots,\thickhat r_m)\in\mathbb R^m_{+}$   is a critical point of the function $\phi$ because   ${\nabla_{r} \,\phi(\thickhat r)=0}$. It is also a saddle point of the analytic continuation of $\phi$ to the interior of the  unit polydisc in $\mathbb C^m$.  We set $r_k=\thickhat r_k$ to make the contours in  \eqref{cauchy} pass through the saddle point of $\phi$.   Let $H_\phi(\thickhat r)$ denote the ${m\times m}$ Hessian matrix of $\phi(r_{})$ 
evaluated at  $\thickhat r$.   The   particle weight $w(x)$ is approximated by $\widetilde w(x)$, where 
\begin{align}
    \label{atlast}
\boxed
{\widetilde w(x)\equiv \frac{e^{\phi(\thickhat r)}}{(2\pi)^{M/2}\sqrt{\det\big[\mathrm{diag}(\thickhat r)^T H_\phi(\thickhat r)\, \mathrm{diag}(\thickhat r)\big]}},
}
\end{align}
where $\mathrm{diag}(\thickhat  r)$ is an ${m\times m}$ diagonal matrix whose diagonal is   $\thickhat r$.  The approximation  depends on  $x$ since, from   \eqref{logPsi},  the   saddle point $\thickhat r = \thickhat r(x)$  is a   function of $x$.

The saddle point $\thickhat r(x)$ for J-PDA  can be computed by a fixed point iteration that  experience shows converges rapidly.  The proof in \cite{FSS} is for  J-PDA, but it  suggests that other  fixed point iterations might be found for other filters.  

The ${m\times m}$ determinant in the saddle point approximation is shown in  \cite{FSS} to be equal to an ${N\times N}$ determinant. The proof uses the Weinstein-Aronszajn identity (a corollary of Sylvester’s
identity). When the number of measurements $m$ is  larger than the number of targets $N$, as is often the case in practice, the computational complexity of a particle weight calculation is the complexity of computing an ${N\times N}$ determinant, which is known to be $O(N^{2.373})$ at most. Neglecting the cost of computing the saddle point,  the complexity of J-PDA is bounded by  $O(c_{\mathrm{SIR}} N^{2.373})$, where $c_{{\mathrm{SIR}}}$ is the number of particles in the filter.      

Several alternative approximations are also mentioned in \cite{FSS}.  For example, by setting the off diagonal elements of  the Hessian matrix to zero, we obtain a mean field approximation to the particle weight, $w(x)$. This approximation avoids computing the full Hessian matrix, and it  avoids computing the determinant.

The minimization problem \eqref{argmnin} has a unique solution for PGFLs that are not polynomials  \cite{FS}. This result is derived from the fact that the PGFLs have only non-negative coefficients in their power series expansions about zero.  The PGFL of J-PDA is not a multivariate polynomial and, therefore,  the solution to \eqref{argmnin} is unique. In this sense, false alarms save the day.  

Error analysis of the saddle point approximation is very difficult in  general.  Consequently, the quality of the   approximation  must be examined on a case by case basis.  Higher order terms in the approximation can be derived, if needed.


\bigskip
\section{Concluding Remarks}
\label{conclude}

The class of PGFL-based   tracking filters is large and diverse, as the  PGFLs gathered together  in this paper show.   PGFLs are remarkably precise expressions derived directly from the probabilistic assumptions that define the  tracking problem.  They completely characterize the problem and provide explicit expressions for many important statistics associated with the filters, as well as approximations to these statistics via the saddle point method.  

The PGFL for  problems with multiple independent labeled targets (or sensors, or both) can often be  turned  into a PGFL for superposed targets simply by making all the indeterminate functions the same.  This changes the product of $N$, say, individual PGFLs into the $N^{\mathrm{th}}$ power of a single PGFL, provided also that the underlying target models are  the same.  This step by itself is important because it can significantly change the computational complexity of the filter. For example, as discussed in Section \ref{JPDA/S}, the labeled J-PDA filter is NP-hard, while the unlabeled U-J-PDA problem is not.  In some problems it  is possible to make $N$ a random integer and thereby find the PGFL of a Bayesian model-order selection problem conditioned on a set of measurements. Perhaps the best known example of this is discussed in Section \ref{CPHDandPHD}, wherein the unlabeled J-PDA filter is turned into the CPHD intensity filter.       

Tracking filters for unlabeled targets on  discrete (e.g., gridded) state  spaces $\mathcal X$ use PGFs and  high order mixed ordinary derivatives.  \cite{2014Streit_FilterOnDiscreteSpaces}. They are of interest in some digital signal processing problems, but they are  outside the scope of the paper.   

Several combinatorially  interesting  statistics  are not  discussed in the paper.  Prominent examples are the pair correlation function, which is useful for understanding  track coalescence in multitarget filters,  and the reduced Palm process, which is useful for sequential  track extraction. They are discussed in    \cite{bogdog}.  Both are computed from mixed first-order derivatives of the PGFL. 

\section*{Acknowledgements}The author thanks the anonymous reviewers of this paper for their helpful  comments and suggestions. The author  thanks Jan Krej\v{c}\'{i} (University of West Bohemia) for his  insightful comments on several sections of the paper.

\bigskip
\appendix[\normalsize{PGFs and PGFLs for Finite Point Processes}]
\label{gfTheory}

The PGF of the random integer ${N\geqslant 0}$ is defined by
\begin{align} 
\label{genPGF}
\mathrm G^N(z)=\textstyle\sum_{n=0}^\infty \,\Pr\{N=n\}\,z^n,
\end{align}
where  $z$ is a complex-valued indeterminate variable. 
If $N$ is Bernoulli distributed, then   ${N=1}$ and  ${N=0}$ with probabilities $p$ and $1-p$, respectively, where  ${p \equiv \Pr\{\text{``success''}\}}$ is the expected value. The  PGF of $N$ is    
\begin{align}
\label{BernoulliPGFAppdx}
    {\mathrm G^{\mathrm {Bern}(p)}(z)=1-p + p\, z}.
\end{align} 
If $N$ is  Poisson   distributed,   then ${\Pr\{N=n\}=e^{-\lambda} \lambda^n/n!}$, where $\lambda\geqslant 0$ is the expected value.  Its  PGF is  
\begin{align}
    \label{PoissonGF}
    \mathrm G^{\mathrm{Pois}(\lambda)}(z)=\sum_{n=0}^\infty \frac{e^{-\lambda} \lambda^n}{n!}\,z^n=\exp(-\lambda+\lambda\,z).
\end{align}
Bernoulli and Poisson distributions are commonly used in tracking problems. Bivariate and multivariate PGFs are defined in a similar way.  The PGF  form of Bayes Theorem  is  a normalized mixed higher order  derivative  \cite{1997JohnsonKotzBalakrishnan}.  


The Bayesian tracking problems cataloged  in this paper are formulated in terms of finite point processes whose PGFLs have  special forms.  When there is only one point process, e.g., a false alarm process, the PGFL  has the form
\eqref{PGFL1}.  This expression is explained  below, but loosely speaking, the PGFL of such a process replaces the indeterminate variable    $z$ in the PGF  \eqref{genPGF} with the  linear functional $\textstyle\int_{\mathcal X} h(x) p(x)\,\mathrm dx$.  Univariate processes of this kind are called i.i.d. point processes \cite{2003DaleyVereJones}. They are reviewed in the first three subsections.  

{{Bivariate}} finite point processes whose joint PGFLs have the special form \eqref{easyBi} are called i.i.d. cluster processes \cite{2003DaleyVereJones}. They are  discussed  in  the last subsection of this appendix.  The  PGFL of the exact Bayesian posterior process for i.i.d. cluster processes is the normalized mixed first-order derivative of the PGFL of  the joint process.  The PGFL form of Bayes Theorem for finite point processes on continuous spaces was first  derived in   \cite{BakIvan} and  appears later in \cite{2007Mahler}.       

This appendix is written in an informal 
style  appropriate to a  tutorial.  A longer  discussion of  multivariate PGFs  in this ``measure-free''  style is available in \cite[Appx. A]{2021ACBook}.  An extended version of the discussion of PGFLs and secular functions in the same style is given in   \cite[Appx. B]{2021ACBook}.  Readers interested in rigorous measure-theoretic discussions will find them in the well-known treatise \cite{2003DaleyVereJones} and the exhaustive collection of  references therein.  

\subsection{Univariate I.I.D. Finite   Point Processes}
\label{iidfpp}

{\it{1) Realizations}} are  multisets\footnote{A multiset is a set that allows multiple instances of each of its elements. The cardinality of a multiset is the sum of the  multiplicities of the distinct elements. According to Knuth \cite[p. 694]{KnuthVol2Ed3}, the concept is very old but the word {\it{multiset}} was coined by N.G. de Bruijn in the 1970s.}  
with a random number, ${N\geqslant 0}$, of i.i.d. sample points in a space $\mathcal X$, where the  sampling distribution  is the same for all values of  $N$. The PGF of $N$ is given by \eqref{genPGF}.  
Points  are   unlabeled, which means  that the realization $\{x_1,\ldots,x_N\}$ is   indistinguishable from a realization that lists the same points  in a different order,  say, reverse order $\{x_{N},\ldots, x_{1}\}$. 

The space $\mathcal X$ is usually Euclidean, but it can also be discrete (e.g., gridded) or  discrete-continuous.  If $\mathcal X$ is discrete, the i.i.d. sampling process can yield repeated samples of the same point with nonzero probability. If $\mathcal X$ is continuous and the sampling distribution $p(x),\,x\in\mathcal X,$ has no point masses, then the  outcomes are sets  with  probability one because i.i.d.  sample points are, in this case,  distinct with probability one. In the tracking literature, a random finite multiset (RFMS) or, less accurately, a random finite set (RFS) is   an i.i.d.  finite point process that is (unnecessarily) restricted to a continuous space $\mathcal X$ with sampling distribution $p(x)$ with no point masses.    Outcomes of an RFMS are always multisets in $\mathcal X$, and they are sets with probability 1. 

The superposition of   several point processes is a point process whose realizations are (multiset) unions of  realizations  of  the constituent processes.  The points in a realization of a superposed process are unlabeled,  that is, it is not known which  constituent process generated which point in the realization.  

The PGFL of a superposed process  is the product of the PGFLs of the constituent precesses --- if the processes are mutually independent.  This  fact is not proved here, but it is important because independence assumptions often hold in tracking applications.  This fact  enables the PGFLs of difficult problems to be derived directly from the   problem statement.   

{\it{ 2) The PGF}} of an i.i.d. process with realizations in a discrete space 
$\mathcal X=\{x_1,\ldots,x_M\}$,  $M\geqslant 1$, has   the form 
\begin{align}
    \label{histogramGFN}
    \Psi(h_{1:M})= \mathrm G^{N}\Big(\textstyle\sum_{m=1}^M \Pr\{x_m\}\,h_m\Big),
\end{align}
where $\Pr\{x_m\}$ is the   probability of sampling $x_m$, and where $h_m$ is the indeterminate variable for the number  $N_m$ of samples of $x_m$. To see this, let ${N=N_1+\cdots+N_M}$ denote the total number of samples in a realization.   
The PGF for a realization with $N=1$ point  is 
\begin{align}
     \label{histogramGF1}
     &\mathrm G\big(h_{1:M}|N=1\big)\nonumber\\
     &\,\,\,=\mathop{\textstyle\sum_{n_1=0}^\infty\cdots \textstyle\sum_{n_M=0}^\infty}_{n_1+\cdots+n_M=1} \Pr\{N_1\!=\!n_1,\ldots,N_M\!=\!n_M\}\, h_1^{n_1}\cdots h_M^{n_M}\nonumber\\
    &\,\,\,={\textstyle\sum_{m=1}^M }\Pr\{x_m\}\,h_m, 
\end{align}
where the first equation defines the multivariate PGF conditioned on $N=1$.  The PGF conditioned on  $N=n$ i.i.d. realizations is 
\begin{align}
    \label{histogramGFn}
    \mathrm G\big(h_{1:M}|N\!=\!n\big)=\Big(\textstyle\sum_{m=1}^M \Pr\{x_m\}\,h_m\Big)^{\!n}.
\end{align}
Multiplying both sides by $\Pr\{N\!=\!n\}$, summing  $n$ from ${0}$ to $\infty$, and using the PGF \eqref{genPGF} of $N$   gives  \eqref{histogramGFN}. 
 
Realizations on discrete spaces are equivalent to histograms.   The probability of a histogram with counts  $(n_1,\ldots,n_M)$ is the mixed derivative of \eqref{histogramGFN} evaluated at  $\mathbf 0_{M}=(0,\ldots,0)\in\mathbb R^M$,
\begin{equation}
    \label{histogramm}
\begin{aligned}
    \Pr\{n_1,\ldots,n_M|n\}&=\frac{1}{n_1!\cdots n_M!} \mathrm D^{(n_1,\ldots,n_M)}\Psi(\mathbf 0_{M})\\
    &=\frac{n!}{n_1!\cdots n_M!} \prod_{m=1}^M \big(\Pr\{x_m\}\big)^{n_m},
\end{aligned}
\end{equation}
where  ${n=n_1+\cdots+n_M}$ and    
\begin{align}
\label{DiffOperator}
    \mathrm D^{(n_1,\ldots,n_M)}\equiv \frac{\partial^n}{\partial h_1^{n_1}\cdots \partial h_M^{n_M}}
\end{align}
is the  mixed partial derivative operator.

{\it{ 3) The PGFL}} of an i.i.d. process on a continuous space \label{continuousSpaces}
is derived by gridding the space, writing the PGF of   an i.i.d. process on the grid cells, and taking the ``small  cell limit'' of the PGF. (Intuitively, the procedure is akin to  deriving an integral as the limit of a Riemann sum.) To keep things simple, we assume the space  $\mathcal X$ is closed, bounded, and convex subset of a Euclidean space.  We partition $\mathcal X$  into a finite number of $M$ non-overlapping  cells, $\{\Delta(m)\}_{m=1}^M$. The grid cells constitute a discrete space.  We choose $x_m$ to be interior to  $\Delta(m)$, and let $\Delta_m$ denote the size of cell $\Delta(m)$. We map the probability vector 
$$
\big(\Pr\{x\in\Delta(1)\},\ldots,\Pr\{x\in\Delta(M)\}\,\big)$$ 
to a step pdf, $p_{\mathrm{step}}(x),\,x\in\mathcal X,$  whose value,  $p_{\mathrm{step}}(x_m)$, at the $m^{\mathrm{th}}$ step is chosen so that 
\begin{align}
    \Pr\{x\in\Delta(m)\}=p_{\mathrm{step}}(x_m) \Delta_m,\quad m=1,\ldots, M.
\end{align}
We  map the indeterminate vector $h_{1:M}$ to a complex-valued step function, $h_{\mathrm{step}}(\cdot)$, whose step values are indeterminates, i.e., 
\begin{align}
    h_{\mathrm{step}}(x)=\sum_{m=1}^M h_m \,\mathcal I\{x\in\Delta(m)\},
\end{align}
where ${\mathcal I\{x\in\Delta(m)\}}=1$ if    ${x\in \Delta(m)}$ and is equal to 0 otherwise. With these mappings, the  PGF \eqref{histogramGFN} takes the form 
\begin{align} \label{ContinuoushistogramGFN} \Psi(h_{\mathrm{step}})= \mathrm G^{N}\Big(\textstyle\sum_{m=1}^M h_{\mathrm{step}}(x_m) \,p_{\mathrm{step}}(x_m)\,\Delta_m\Big).
\end{align}
The argument of $\mathrm G^{N}$ resembles a Riemann sum and, instinctively, we want to take the limit  as $M\rightarrow \infty$. Since  $ \mathrm G^N(\cdot)$ is analytic,  we can move the limit into the argument  if the magnitude of $h_{\mathrm{step}}(\cdot)$ is no larger than one.  If ({\it{i}}) the step sizes $\Delta_m$ go to zero uniformly,  ({\it{ii}}) $p_{\mathrm{step}}(\cdot)$  converges to a locally integrable function $p(\cdot)$, and ({\it{iii}}) $h_{\mathrm{step}}(\cdot)$  converges to a locally integrable function $h(\cdot)$, then  the ``small cell limit'' of the PGF as $M\rightarrow\infty$ is the PGFL of the i.i.d. finite point process, 
\begin{align}
    \label{PGFL1}
\Psi(h)= \mathrm G^{{ N}}\Big(\textstyle\int_{\mathcal X} h(x) p(x)\,\mathrm dx\Big).
\end{align}
This is the right result, but the derivation is problematic because it raises the question, ``Is the limit of {\it{indeterminate}} step functions the same as the limit of step functions?''   If  not, then \eqref{PGFL1} is not a well defined integral. 

We avoid the question by restricting the integral to locally integrable functions  $h(\cdot)$ and justify calling  \eqref{PGFL1} the PGFL of the process by showing how to use it to recover the event probabilities.  Setting $h(x)\equiv z$ in \eqref{PGFL1} reduces the PGFL to  the PGF \eqref{genPGF} of $N$.  It remains to show how to recover  the  probability of realizations  with   $N=n$ points.  
Let $\mathbf x_{1:n}=\{x_1,\ldots,x_n\}\subset \mathcal X$  be one such  realization. We  define, for $x\in\mathcal X$, the delta operator $\delta_x(\cdot)$ on continuous integrable functions ${f:\mathcal X\rightarrow \mathbb C}$ by the property ${\int_{\mathcal X} f(x')\delta_x(x')\,\mathrm dx'=f(x)}$.     Substituting the weighted delta train 
\begin{align}
    \label{deltatrain}
    h_{\mathrm{sec}}(x)=\textstyle\sum_{j=1}^n h_j\,\delta_{x_j}(x),\quad x\in\mathcal X,
\end{align}
into \eqref{PGFL1} yields  a function that we call the ``secular'' function: 
\begin{align}
    \label{secular}
    \Psi_{\mathrm{sec}}(h_{1:n})\equiv  \mathrm G^{{ N}}\big(\textstyle\sum_{j=1}^n h_j p(x_j)\big).
\end{align}
It is a complex-valued analytic function in each variable $h_j$ separately, and (by Hartogs's Theorem) it is jointly analytic in all $n$ variables $h_{1:n}$.  Each point in $\mathbf x_{1:n}$ occurs  once (with probability one since  $p(\cdot)$ is locally integrable and has no point masses). Evaluating the mixed  first-order derivative  of  \eqref{secular} at $h_{1:n}=\mathbf 0_n$ gives (cf. \eqref{histogramm} - \eqref{DiffOperator})
\begin{subequations}
    \label{connection}
\begin{align}
    \label{simpleA}p(\mathbf x_{1:n})&=   \mathrm D^{(\mathbf 1_{n})}\Psi_{\mathrm{sec}}(\mathbf 0_{n})\\
    &=n! \Pr\{n\} \prod_{j=1}^n p(x_j),
\end{align}
\end{subequations}
where ${\mathbf 1_{n}=(1,\ldots,1)\in\mathbb R^n}$. The result \eqref{connection}  is the probability density that a realization has $n$ distinct points and that these points are   $\mathbf x_{1:n}$. The  procedure outlined here is  mathematically rigorous \cite{2014Streit}.  It is also discussed in \cite{2021ACBook}. 

To compute the intensity function $\Lambda(x_0)$ at  a point $x_0\in\mathcal X$,  substitute  $h_{\mathrm{sec}}(x)=1+h_0 \delta_{x_0}(x)$  into \eqref{PGFL1}, where $h_0$ is a complex-valued indeterminate variable, and let $\Psi_{\mathrm {sec}}(h_0)\equiv \Psi(h_{\mathrm {sec}})= \mathrm G^{ N}\big(1+ h_0\,  p(x_0)\big)$.  The first derivative at $h_0=0$ is 
\begin{equation}
\begin{aligned}
    \label{PGFL12}
\Lambda(x_0)&=\mathrm D^{(1)}\Psi_{\mathrm {sec}}(0)\\
&\equiv \left.\frac{\mathrm d}{\mathrm dh_0}\right|_{h_0=0}\Psi_{\mathrm {sec}}(h_0)=E[N]\, p(x_0).
\end{aligned}
\end{equation}
The  intensity function is the product of $E[N]$, the  expected number of points in a realization, and  $p(x_0)$, the pdf of  the points in a realization.  

The PGF of $N$ is derived from the  PGFL \eqref{PGFL1} by  setting $h(x)= z$, where $z$ is a complex-valued indeterminate variable. Define  $\Psi(z)\equiv \Psi(h)\vert _{h(\cdot)=z}$. Then
\begin{equation}
\begin{aligned}
    \label{pdfOfN}
    \Pr\{N=n\}&=\frac{1}{n!}\mathrm D^{(n)}\Psi_{\mathrm {sec}}(0)\\
    &\equiv  \left.\frac{1}{n!}\frac{\mathrm d^n}{\mathrm dz^n}\right |_{z=0}\Psi(z),\quad n \geqslant 0,
\end{aligned}
\end{equation}
is the PGF of $N$.  

PGFLs assign zero probability to  realizations that are disallowed, or infeasible.  Consider, for example,  a process whose realizations  cannot have  $n_0$ points and  a point set $S$ that has $n_0$ points.  The weighted delta train for $S$  has  $n_0$ terms, and substituting it  into the PGFL gives a secular function with $n_0$ variables. If the mixed first-order partial derivative of this  function evaluated at zero is {\it{not}}  zero, then  $S$ is non-zero probability, contradicting the assumption that the process cannot have realizations of size $n_0$.  Thus, if the  PGFL  is correctly formulated, the mixed first-order derivative is zero.     

\subsection{Bivariate Cluster Processes and  Bayes Theorem}
\label{pointprocesses} 
A  bivariate finite point process consists of two finite point processes, one on the space $\mathcal X$ and the other on the space $\mathcal Y$.  For simplicity, we  assume  that both spaces are continuous. We will refer to  $\mathcal X$ as the target state space and $\mathcal Y$ as the measurement space.  For tracking applications, we  consider  a subclass of bivariate finite point processes called i.i.d. cluster processes \cite[Sect. 6.3]{2003DaleyVereJones}.  We  will call the  process on $\mathcal X$   the target process and the process on $\mathcal Y$  the measurement process. 

The target process is assumed to be an i.i.d. point process. Its realizations have a random number ${N\geqslant 0}$ of  samples, each of which is i.i.d.   according to  a  pdf $p(x)$ that has no  point masses.  The PGF of $N$ is defined by \eqref{genPGF}, where $z$ is the complex-valued indeterminate  variable for targets.  

Let  ${\mathbf{x}_{N}\subset\mathcal X}$ be a realization of the target process.  The points in $\mathbf{x}_{N}$ are distinct with probability 1 because $p(x)$ has no point masses. The  measurement process is defined to be the superposition of $N$ i.i.d. point processes, one for each   point in   ${\mathbf{x}_{N}}$. Realizations of  the  measurement process conditioned on ${x\in\mathbf{x}_N}$   have a random  number ${M|x\geqslant 0}$ of i.i.d. sample points in $
\mathcal Y$ distributed according to the  conditional pdf $p(y|x)$. These points are distinct with probability 1 because $p(y|x)$ has no  point masses. The PGF of $M|x$ is 
\begin{align}
    \mathrm G^{ M|x}(w)=\textstyle\sum_{m= 0}^\infty \Pr\{M\!=\!m\,|\,x\} \,w^m,
\end{align}
where $w$ is the complex-valued indeterminate variable for measurements.  Applying the univariate  result \eqref{PGFL1} to the  measurement process conditional on $x$ gives the  PGFL of the process as 
\begin{align}
\label{clusterConditional}
     \mathrm G^{ M|x}\Big(\!\textstyle\int_{\mathcal X} g(y) p(y|x)\,\mathrm dy\Big).
\end{align}
The PGFL of the  joint target-measurement process for $N=1$ is  
$$
\int_{\mathcal X}  h(x) p(x) \, \mathrm G^{ M|x}\Big(\!\textstyle\int_{\mathcal X} g(y) p(y|x)\,\mathrm dy\Big)\,\mathrm dx, 
$$
where  $h(\cdot)$ is the  complex-valued indeterminate function for targets.   There are  $N$ i.i.d.  processes, so    
\begin{align}
\label{easyBi}
\Psi(h,g)= \mathrm G^{ N}\!\left(\int_{\mathcal X}  h(x) p(x) \, \mathrm G^{ M|x}\Big(\!\textstyle\int_{\mathcal X} g(y) p(y|x)\,\mathrm dy\Big)\,\mathrm dx\right)
\end{align}
is the  PGFL of the i.i.d.  cluster process model of  the bivariate   target-measurement process.

The bivariate secular function  is found by substituting two delta trains into \eqref{easyBi}. Neither $p(x)$ nor $p(y|x)$ have point masses, so the delta  train for $h$ is    \eqref{deltatrain} and  the delta train for $g$ is 
\begin{align}
    \label{xxxx}
    g_{\mathrm{sec}}(y)=\textstyle\sum_{k=1}^m g_k\,\delta_{y_k}(y),\quad y\in\mathcal Y,
\end{align}
where the coefficients  $g_{1:m}$ are  complex-valued indeterminate variables.  The  secular function with respect to both $h$ and $g$ is 
\begin{equation} \label{clusterppsecular}
\begin{split}
\Psi_{\mathrm{sec}}&(h_{1:n},g_{1:m})\equiv \Psi(h_{\mathrm{sec}},g_{\mathrm{sec}})\\
&= \mathrm G^{ N}\!\left(\sum_{j=1}^n h_j\,p(x_j)\, \mathrm G^{ M|x_j}\bigg(\sum_{k=1}^m g_k \,p(y_k|x_j)\bigg)\right).
\end{split}
\end{equation}
This function is jointly analytic in the complex-valued indeterminate variables $h_{1:n}$ and $g_{1:m}$.    Paralleling the notation of  \eqref{simpleA}, the joint pdf of $(\mathbf{x}_{1:n},\mathbf{y}_{1:m})$  is the mixed first-order derivative of the joint secular function with respect to all $n+m$ variables $h_{1:n}$ and $g_{1:m}$ evaluated at zero,  
\begin{align}
    \label{doubel}
    p(\mathbf{x}_{1:n},\mathbf{y}_{m})=\mathrm D^{(\mathbf 1_{n},{\mathbf 1}_{m})}\Psi_{\mathrm{sec}}(\mathbf 0_{n},\mathbf 0_{m}).
\end{align}
The probability  distribution of the joint finite point process is encoded in the first-order derivatives of the secular function. 

The Bayes posterior process on $\mathcal X$ is the finite point process that results from conditioning on the realization $\mathbf{y}_{1:m}$ in $\mathcal Y$.  Its PGFL is the normalized mixed first-order (partial) derivative of the  secular function for $g$ with respect to the  variables $g_{1:m}$  evaluated at $\mathbf 0_{m}$,  
\begin{equation}
\boxed{  
\begin{aligned}
    \label{BayesIsBeautiful}
    \Psi^{\mathrm{Bayes}}(h|\mathbf{y}_{1:m})=\frac{\mathrm D^{(0,\mathbf 1_m)}\Psi_{\mathrm{sec}}^{}(h,\mathbf 0_{m})}{\mathrm D^{(0,\mathbf 1_m)}\Psi_{\mathrm{sec}}(1,\mathbf 0_{m})}\,,
\end{aligned}
}
\end{equation}
where  the secular  function with respect to $g$ alone is    
\begin{align}
    \label{BayesSec}
    \!\Psi_{\mathrm{sec}}(h,g_{1:m})\!\equiv\!  \mathrm G^{ N}\!\!\left(\int_{\mathcal X} \!\!h(x) p(x) \,\mathrm G^{ M|x}\bigg(\sum_{k=1}^m g_k \,p(y_k|x)\!\bigg)\mathrm dx \!\right)\!\!.
\end{align}
The  posterior intensity of $\Psi^{\mathrm{Bayes}}(h|\mathbf{y}_{1:m})$ is computed   using \eqref{PGFL12}. When the only nonzero probability  realizations   have exactly one target, then the posterior process is equivalent to a random variable and the intensity function is the Bayes posterior pdf.  

An example of the general  procedure is given in detail for the PGFL of the PDA filter  in Section \ref{PDAexample}.  See Eqns. \eqref{PDA}-\eqref{PDApdfExample}.  

\bigskip

\end{document}